\documentclass[fleqn,usenatbib]{mnras}
\usepackage{newtxtext,newtxmath}

\usepackage[T1]{fontenc}
\usepackage{ae,aecompl}
\usepackage[para,online,flushleft]{threeparttable}

\usepackage{graphicx}	
\usepackage{amsmath}	
\usepackage{amssymb}	
\usepackage[flushleft]{threeparttable}
\usepackage{float}
\usepackage[caption = false]{subfig}
\usepackage{enumitem}
\usepackage{comment}
\usepackage{colortbl}
\usepackage{dblfloatfix}    
\usepackage{multirow}
\usepackage{booktabs}

\newcommand{\ra}[1]{\renewcommand{\arraystretch}{#1}}
\newcommand{\xmm}{{\sl XMM-Newton }}
\makeatletter
\newcommand\footnoteref[1]{\protected@xdef\@thefnmark{\ref{#1}}\@footnotemark}
\makeatother

\title[Multipolar Magnetic Field in PSR J0108--1431]{Evaluating the Evidence of Multipolar Surface Magnetic Field in PSR J0108--1431}

\author[Arumugasamy \& Mitra]{
Prakash Arumugasamy,$^{1}$\thanks{E-mail: prakash@ncra.tifr.res.in}
Dipanjan Mitra,$^{1,2}$
\\
$^{1}$National Centre for Radio Astrophysics, Tata Institute for Fundamental Research, Post Bag 3, Ganeshkhind, Pune 411007, INDIA\\
$^{2}$Janusz Gil Institute of Astronomy, University of Zielona G\'ora, ul. Szafrana 2, 65-516 Zielona G\'ora, Poland
}

\date{Accepted XXX. Received YYY; in original form ZZZ}

\pubyear{2019}

\begin{document}
\label{firstpage}
\pagerange{\pageref{firstpage}--\pageref{lastpage}}
\maketitle

\begin{abstract}
PSR J0108--1431 is an old pulsar where the X-ray emission is expected to have a thermal component from the polar cap and a non-thermal component from the magnetosphere.
Although the phase-integrated spectra are fit best with a single non-thermal component modeled with a power-law (PL) of photon index $\Gamma=2.9$, the X-ray pulse profiles do show the presence of phase-separated thermal and non-thermal components.
The spectrum extracted from half the rotational phase away from the X-ray peak fits well with either a single blackbody (BB) or a neutron star atmosphere (NA) model, whereas, the spectrum from the rest of the phase range is dominated by a PL.
From Bayesian analysis, the estimated BB area is smaller than the expected polar cap area for a dipolar magnetic field with a probability of 86\% whereas the area estimate from the NA model is larger with a probability of  80\%.
Due to the ambiguity in the thermal emission model, the polar cap area cannot be reliably estimated and hence cannot be used to understand the nature of the surface magnetic field.
Instead, we can infer the presence of multipolar magnetic field from the misalignment between the pulsar's thermal X-ray peak and the radio emission peak.
For J0108--1431, we estimated a phase-offset $\Delta\phi > 0.1$ between the thermal polar cap emission peak and the radio emission peak and argue that this is best explained by the presence of a multipolar surface magnetic field.
\end{abstract}


\section{Introduction}\label{sec:intro}
Rotation-powered pulsars (RPPs) are rapidly rotating, highly magnetized neutron stars (NSs) that have active magnetospheres, where efficient plasma production and acceleration processes can occur.
Due to the rapid rotation, the magnetosphere is divided into regions of open and closed magnetic field lines depending on whether or not they extend beyond the light cylinder --- an imaginary cylinder around the rotation axis where the corotation speed equals the speed of light.
In the standard model of pulsars, a strong co-rotating electric field populates the NS exterior by pulling out charges from the NS surface \citep{GoldreichJulian1969} and/or creating electron-positron pairs through magnetic pair production \citep{Sturrock1971}.
The charged particles achieve a force-free (FF) configuration due to the strong electromagnetic forces that vastly exceed the inertial and dissipative forces.
Under such conditions, the minimum charge needed to screen the electric field is the Goldreich-Julian number density given by $n_{\rm GJ} = \Omega\cdot B /2\,\pi\,e\,c$, where, $B$ is the magnetic field, $\Omega = 2\,\pi/P$ ($P$ is pulsar period), $e$ is the electric charge, and $c$ is the speed of light in vacuum.
The closed field line region is filled with dense plasma which screens the electric field and co-rotates with the star.
The FF conditions also provide prescriptions for the charge density and current density in the open field line regions.
However, the FF conditions need to be relaxed in the open field line regions to explain observations of energetic particle winds and high-energy radiation that are powered by the pulsar's spin-down.

{}The evidence for particle wind comes from observations of bright wind nebulae around young pulsars \citep{Kargaltsev2015}, and modeling such wind nebulae requires high pair multiplicity factors, $\kappa \lesssim 10^5 n_{\rm GJ}$ as suggested by the pair cascade models (see e.g. \citealt{deJager2007,Timokhin2019}).
{}The magnetospheric origin of the electromagnetic radiation is apparent from the pulsed and co-rotating radiation patterns observed from the radio to the $\gamma$-rays in several pulsars.
The open field lines regions, hence, require ``gap'' regions where copious pair creation, acceleration, and cascade takes place, which leads to a highly-relativistic flow of charges as particle winds and the electromagnetic broadband radiation.
Although the modeling of FF magnetosphere has been successful in providing global solutions for the magnetic field structure and the plasma dynamics in the pulsar magnetosphere (see for e.g. \citealt{Michel1999, Contopoulos1999, Spitkovsky2011, Petri2016}), it is yet to incorporate self-consistent particle acceleration and radiation models (\citealt{Cerutti2017}; and references therein).
Observations that constrain the pair cascade gap regions and the local magnetic field geometry are crucial to determine regions where the FF conditions are broken.

To obtain such observational constraints on the magnetic field structure, in this work we will focus on emission from normal RPPs with periods longer than 100 ms.
Observations of a few pulsars in this category show both incoherent X-ray and coherent radio emission.
The X-ray spectra are usually modeled to have a combination of thermal and non-thermal emission components.
The thermal emission is thought to originate from the polar cap\footnote{Younger pulsars, with characteristic age $\tau\lesssim1$ Myr, have an additional X-ray emission component from the entire stellar surface as they lose their heat of formation.}.
The physical location of the non-thermal emission, on the other hand, is uncertain and various models suggest that the emission can originate from different parts of the magnetosphere (\citealt{Becker2009,Cerutti2017}).
The coherent radio emission has been constrained by observations to originate from regions below 10\% of the light cylinder (see for e.g. \citet{Mitra2017JApA} and references therein).
These features of RPP emission are best explained by  \cite{RudermanSutherland1975}, RS model of pulsar emission, where it was postulated that a charge-starved ``inner vacuum gap'' (IVG) region exists just above the pulsar polar cap (PC) where pair plasma can be created, accelerated, and cascaded.
{}The efficiency of pair creation depends on the curvature of the magnetic field lines in the IVG, and to get $\kappa \sim 10^5$, the dipolar magnetic field radius of curvature is too large and hence inadequate \citep{Timokhin2019}.
Instead, the presence of strong multipolar magnetic field lines with smaller radii of curvature is more efficient at increasing the pair multiplicity.

{}The electric field in the IVG accelerates the charged particles of one sign upwards along the open-field lines and the oppositely charged ones downwards onto the polar cap \citep{RudermanSutherland1975}.
{}The particle bombardment on the polar cap heats up the surface to temperatures $kT\gtrsim0.1$ keV (\citealt{Ogelman1991,Cheng1980,Harding2001}).
This has been observed as thermal emission from emitting regions of sizes much smaller than the NS radius (see e.g. \citealt{Pavlov2002,Geppert2017,Rigoselli2018} and references therein).
{}This emission flux depends on the energy dumped by the accelerated charged particles and the PC area.
{}For a global dipolar magnetic field, the PC area is calculated by finding the locus of the last closed field lines on the NS surface.
{}A multipolar magnetic field geometry has dipolar and other higher order contributions near the surface while maintaining a purely dipolar structure away from the surface 
\citep{Gil2002}.
{}Due to magnetic flux conservation, such multipolar magnetic field structures are expected to have smaller PC area than the PC area for a pure dipole.
{}Since we detect X-ray thermal emission from localized hot-spots (presumably the PCs) in some RPPs, constraining their emission size can provide evidence for/against a purely dipolar field.

{}In the case of a blackbody, the emission is isotropic and the flux is directly proportional to the emitting area.
The isotropic nature of the BB at all energies produces pulse profiles independent of energy and dependent only on the geometry and inclination of the emitting region.
Hence, measures of profile shape, such as pulsed fraction -- the ratio of counts in the pulse over the total counts, are expected to remain constant over energy for a BB.
{}Once the BB nature of the PC emission is established, its area is estimated from fitting a BB model component to the observed spectrum or by modeling the pulse profile using geometric modulation of a hot spot on a rotating sphere (\citealt{Pechenick1983, Page1995}).
{}The emission area can then, in principle, be compared with the theoretical PC area for a magnetic dipole to establish the presence of a multipolar magnetic field geometry.

{}In reality, however, the emission from the PC can deviate from a BB due to a variety of effects.
{}The NS crust could have anisotropy in thermal conductivity due to the strong magnetic fields that constrain the motion of the electrons (\citealt{Pons2009}; \citealt{Reisenegger2009}, \citealt{Geppert2004}).
{}The star could have a thin atmosphere on the surface, containing elements with energy levels modified by the strong magnetic field, which reprocesses the BB emission (\citealt{Zavlin2002}; \citealt{Pavlov1978}).
{}The strong backflow of particles can modify the thermal emission through inverse Compton scattering (\citealt{Kardashev1984}, \citealt{Sturner1995}, \citealt{Baring2011}).
{}In such cases, determining the emission process and estimating the polar cap area are not straightforward and the solutions may not be unique. In this paper we will often single out BB from thermal emission models where the former is an un-modified blackbody emission whereas the latter, thermal emission, can be either a BB or BB modified by one of the above-mentioned processes.

{}The PC area estimation procedure outlined above to check for multipolar magnetic fields requires definite proof that the PC emission is a BB.
{}However, there is no definitive proof uniquely identifying the thermal emission from pulsars as BB.
{}Phase-integrated spectra show thermal emission as one of its components which are routinely fit with a BB model.
{}But, with current high signal-to-noise (S/N) spectra, the fits from neutron star atmosphere models are statistically indistinguishable from a BB fit \citep{Bogdanov2013}.
{}Rather, pulse profiles in the energy range where thermal emission is dominant are often show large pulsed fractions and modeling shows deviations from the BB profiles \citep{Bogdanov2013}.
{}Due to low statistics of most X-ray detected RPPs, tests of isotropic emission (non-evolution of the pulsed fraction with energy) are inconclusive (e.g., \citealt{Hermsen2018}).
{}Hence, we have not yet identified a single RPP with definite proof of BB emission component from its PC.

{}Our goal is to obtain robust evidence for multipolar magnetic fields by performing a uniform analysis of thermal emission from non-recycled RPPs.
{}There have been attempts to find evidence of multipolar magnetic fields by estimating the PC areas (\citealt{Geppert2017}; \citealt{Rigoselli2018}).
{}However, these works rely on a diverse body of literature, which suffers from one or more of a number of problems (See Section \ref{sec:discussion} for a detailed critique).
{}Foremost of all, convincing evidence for BB emission, even in some cases the presence of thermal components, is not provided.
{}Further, the confidence limits on the BB area obtained from spectral fitting are often underestimated due to the fixing of secondary model components or some of the parameters such as distance and NS radius.
{}We would like to address these issues by attempting a uniform statistical analysis of all thermally emitting, non-recycled RPPs, and determine whether unambiguous proof of BB PC emission can be obtained with the currently available data.
{}The analyses and tests will be customized to characterize the thermal emission from RPPs and estimate the emitting region areas where possible.

In cases where the thermal emission cannot be uniquely modeled, we will find evidence for/against a purely dipolar magnetic field by estimating the offsets between the thermal X-ray and radio peaks (X-R offsets).
The constraint on the radio emission heights ($\lesssim 10\%$ of light-cylinder) and the success of the rotating vector model (RVM; \citealt{Radhakrishnan1969}), for slowly rotating RPPs, is consistent with a dipolar magnetic field structure at the radio emission region.
A star-centered, global dipolar magnetic field geometry will hence show X-ray thermal PC emission peak coincident with the radio peak, after accounting for the effects of aberration and retardation.
Alternatively, measurements of significant X-R offsets will provide qualitative evidence for multipolar magnetic field structure close to the NS surface (see e.g., \citealt{Gil2002, Melikidze2009, Szary2015}).

{}To begin with, in this work, we study the pulsar PSR J0108--1431 (J0108 hereafter), an old RPP where an earlier analysis had shown the possibility of a thermal emission component in addition to a dominant non-thermal component (\citealt{Posselt2012}; BP2012 hereafter).
{}It is representative of a class of old and cooled NSs with characteristic ages $\tau_c\gtrsim1$ Myr (e.g., PSR B1929+10; \citealt{Becker2006}, PSR B0950+08; \citealt{Pavlov2017}, and PSR B1133+16; \citealt{Szary2017}), where the thermal emission from the bulk of the surface is not seen in the X-ray band.
{}The overall X-ray emission (thermal PC and non-thermal) gets fainter with age as the spin-down power reduces \citep{LiLuLi2008}.
{}Much of the analyses in these systems are restricted to obtaining the best S/N pulsations and spectra and testing simple emission models with few parameters.
{}Through spectral and timing analysis, we attempt to find stronger evidence for thermal PC emission from the pulsar and obtain conservative estimates for the model parameters.
{}Then we use the thermal PC emission area and phase of peak flux to evaluate the evidence for a multipolar surface magnetic field.

{}PSR J0108--1431 \citep{Tauris1994} is a slow-rotating ($P=0.808$ s) pulsar with a characteristic age $\tau_c=196$ Myr, at a parallax measured distance of $210_{-50}^{+90}$ pc (Table \ref{tbl-1_summary}; \citealt{Deller2009,Verbiest2012}).
{}It has a spin-down power ($\dot{E}$) of $5\times10^{30}$ erg s$^{-1}$ and an estimated surface magnetic field ($B_{\rm surf}$) of $2.3\times10^{11}$ erg s$^{-1}$, assuming the power loss is entirely via magnetic dipole radiation.
{}It was detected in the X-rays with {\sl Chandra} \citep{Pavlov2009} and later re-observed with \xmm (BP2012).
{}The pulsar showed a relatively soft phase-integrated spectrum well-fit with a PL of photon index $\Gamma\approx3$ in both these data.
{}BP2012 showed that a BB component could be fit only if the PL photon index is fixed at 2.
{}Although X-ray pulsations were detected in the \xmm data, a phase-resolved spectral analysis was not attempted due to low counts.

We re-analyze the data using improved energy-dependent events extraction to obtain pulse profiles with higher pulsations significance.
This has allowed us to select the optimal energy range where the thermal emission can be better distinguished from the non-thermal emission.
Hence, we were able to detect the thermal emission with higher significance and determine reliably the phase at which the emission peaks.

\begin{table}
\ra{1.1}
\centering
\caption{Pulsar J0108--1431 Parameters Summary.}
\label{tbl-1_summary}
\begin{threeparttable}
\begin{tabular}{ll}
\hline\hline
Parameter & Value \\
\hline
Right ascension (J2000){\def\hfill{\hskip 10pt plus 1fill}\dotfill}	&	$01^{\rm h} 08^{\rm m} 08\fs347016(88)$ \\
Declination (J2000) \dotfill	            &     $-14^\circ 31\arcmin 50\farcs1871(11)$ \\
Position epoch (MJD) \dotfill		        &	54100	\\
Galactic longitude/latitude ($l/b$)\dotfill	&	$9\fdg83$ / $-20\fdg06$	\\
Period ($P$) \dotfill						&	0.807564614019(20) s	\\
Period derivative ($\dot{P}$) \dotfill		&	$7.704(12) \times 10^{-17}\;{\rm s\;s}^{-1}$	\\
Frequency ($\nu$) \dotfill					&	1.23829100810(3) Hz	\\
Frequency derivative ($\dot\nu$) \dotfill	&	$-1.1813(18) \times 10^{-16}\;{\rm s}^{-2}$	\\
Epoch of timing solution (MJD) \dotfill		&	50889	\\
Dispersion measure (DM)\dotfill				&	0.23 cm$^{-3}$ pc	\\
Distance\tnote{a} \phantom{"}($d$)\dotfill		&	$210_{-50}^{+90}$ pc	\\
Characteristic age ($\tau_c$) \dotfill		&	$196$ Myr	\\
Spin-down power ($\dot{E}$) \dotfill		&	$4.88 \times 10^{30}$ erg ${\rm s}^{-1}$	\\
Surface magnetic field ($B_{\rm surf}$) \dotfill	&	$2.32 \times 10^{11}$ G	\\
\hline
\end{tabular}
  \begin{tablenotes}
    \item[a] The parallax distance was measured by \cite{Deller2009} and corrected for Lutz-Kelkar bias by \cite{Verbiest2012}.
    \item The parameters are taken from the ATNF catalog \citep{Manchester2005}.
  \end{tablenotes}
\end{threeparttable}
\end{table}

\section{Observations and Data Analysis} \label{sec:dataAnalysis}

We use the archival \xmm data of PSR J0108--1431 from June 15, 2011 (MJD 55727).
The European Photon Imaging Cameras (EPIC) were operated in full-frame mode during the continuous 127 ks exposure.
We mainly work with the EPIC-pn (Struder et al. 2001) data which has a 73.4 ms (less than pulsar period $P=0.808$ s) time-resolution for timing and phase-resolved analyses.
The EPIC-MOS detectors, operated in full-frame mode, have a time-resolution of 2.6 s (greater than the pulsar period) and hence, were used only for phase-integrated spectral analysis.
The data reduction was performed with \xmm Science Analysis Software (SAS) ver. 16.1.0 (Gabriel et al. 2004), applying standard data reduction and events filtering and extraction tasks.

\begin{figure}
\centering
\includegraphics[width=0.47\textwidth]{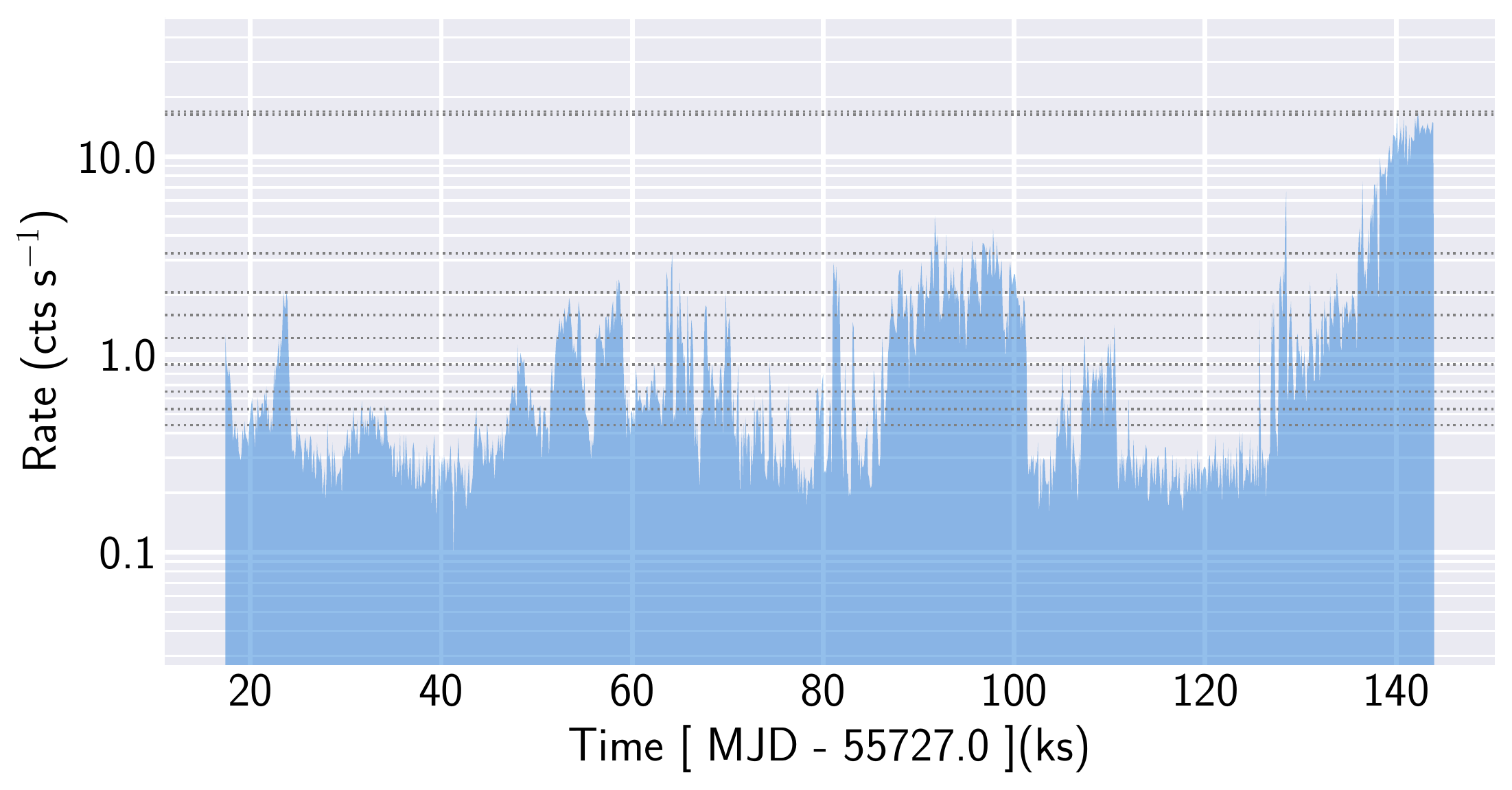}
\caption{\label{fig:fig2_flaring}Background flaring light curve obtained from EPIC-pn (full-frame region) for the full observation duration using events with energies $>10$\,keV.
The dotted lines show the various flaring count-rate cut-offs used to find optimal good time intervals for high S/N events extraction from the pulsar.
}
\end{figure}

We identify time-intervals that are affected by soft proton flaring background by extracting a count rate light curve of single pixel events (from the entire CCD) with energies greater than 10 keV.
At these high energies, the individual source contributions are significantly reduced or negligible and the flaring background contribution dominates.
Soft proton flaring background had affected $\sim 30\%$ of the exposure with count rates exceeding the quiescent level by over $3\sigma$ and approximately $5\%$ with count rates exceeding the quiescent level by $10\sigma$ (Figure \ref{fig:fig2_flaring}).
The good time intervals (GTIs) are obtained by filtering out time ranges when the flaring count rate exceeds a chosen threshold.

\begin{figure*}
\captionsetup[subfigure]{labelformat=parens, labelsep=space}
\subfloat[]{\label{fig:J0108Ia}\includegraphics[width = 0.33\textwidth]{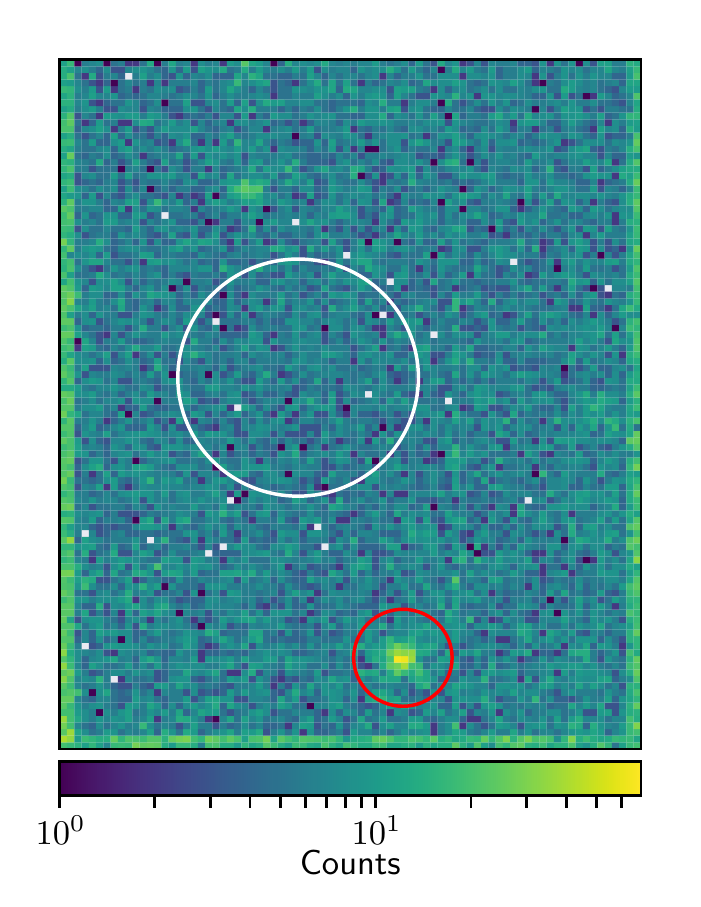}}
\subfloat[]{\label{fig:J0108Ib}\includegraphics[width = 0.33\textwidth]{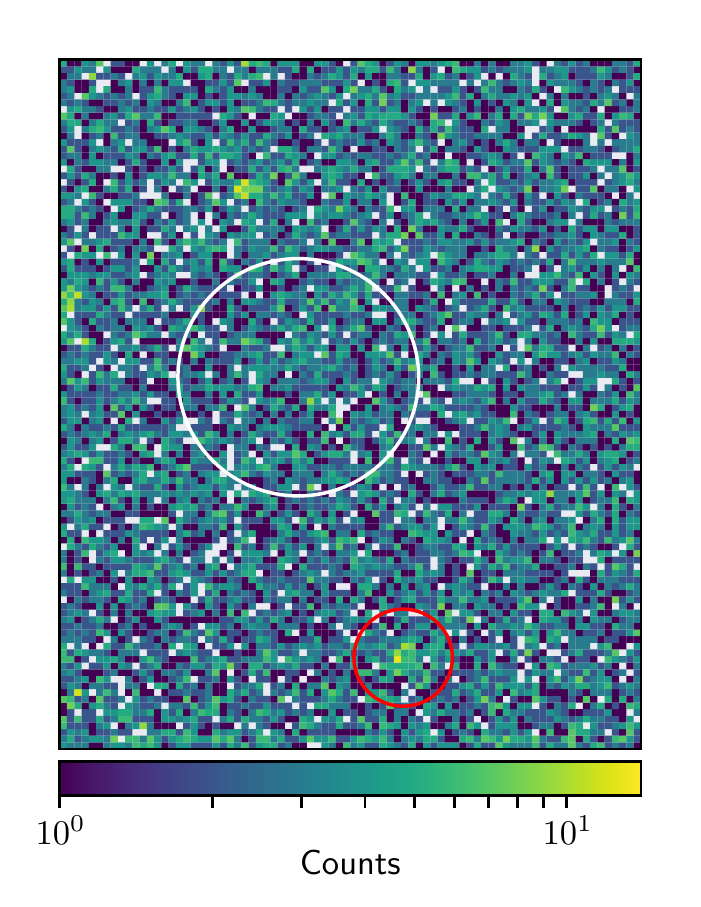}}
\subfloat[]{\label{fig:J0108Ic}\includegraphics[width = 0.33\textwidth]{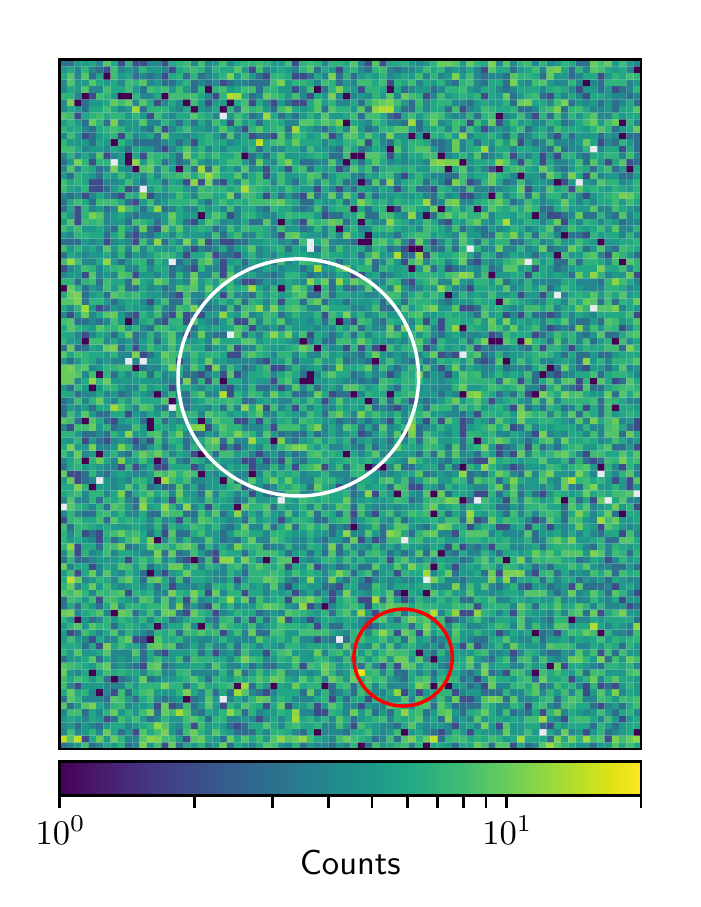}}
\caption{Energy-resolved count maps in the ranges (a) $0.15-2$\,keV, (b) $1-2$\,keV, and (c) $2-5$\,keV from EPIC-pn chip \#4 which contains the source (red circles).
The background regions used for S/N calculations and extracting background spectra are shown in white.
The locations with zero counts are outside the counts color range and hence assigned white.
The images show weak detection between $1-2$\,keV and no detection above 2\,keV in the 127\,ks exposure.}
\label{fig:J0108Simages}
\end{figure*}

The source is detected only in the $0.15-2$ keV energy range with weak detection between $1-2$ keV and no significant contribution above 2 keV (Figure \ref{fig:J0108Simages}).
The sensitivity of the EPIC-pn detector is energy dependent and for a typical pulsar spectrum, the count rates decrease steeply with energy.
This leads to non-uniform signal-to-noise (S/N) with energy for events extracted from a constant size region.
In order to obtain a high S/N events extraction for timing and spectral analysis, we optimize events extraction over the three parameters -- energy range, GTIs, and extraction radius.

To extract an optimal number of photon events at all energies, we determine the S/N independently for the energy bins: $150-300$ eV (first bin), 100 eV bins in the $300-1000$ eV range, 1 keV bins in the $1-7$ keV range, and a $7-10$ keV bin (last bin).
This selection of energy bins is arbitrary, but follows the condition that the bin widths exceed the detector's minimum energy resolution (by a factor $\geq 2$ below 1 keV and $\sim 7-20$ above 1 keV).
Large bins at high energies are preferred due to rapidly decreasing source counts with energy.
In these energy ranges, we determine S/N values over a range of GTI selections and extraction region radii.
The $55\arcsec$ radius circular region used for background events extraction is shown in Figure \ref{fig:J0108Simages}.
Ten sets of GTIs are obtained by choosing flaring count-rate cut-offs between the $3\sigma$ limit over the quiescent level and the highest count rate (Figure \ref{fig:fig2_flaring}) while ensuring a linear increase in total exposure time.
The radii of source extraction apertures range from $5\arcsec$ to $22\farcs5$ in increments of $2\farcs5$.
The center of the extraction circle is chosen by taking the median physical X and Y positions from the events extracted from an initially provided $22\farcs5$ radius extraction region centered approximately on the source.

We obtain optimal parameters for the highest S/N source events extraction from the matrix of S/N values calculated over the range of GTIs and extraction radii for each energy bin.
For J0108, the method did not benefit the spectral analyses and hence the spectra were extracted following the standard procedure using a single GTI selection and extraction region over the full energy range.
But, as shown below, we achieved high significance pulsations not only in the full $0.15-2$ keV range but also in the two energy ranges below and above 0.7 keV.
As a result, we obtained reliable profiles and this prompted further phase-resolved spectral analyses.

\section{Phase-Integrated Emission}

\begin{figure*}
\captionsetup[subfigure]{labelformat=parens}
\subfloat[]{\label{fig:NewNSchematic}\includegraphics[width = 0.33\textwidth]{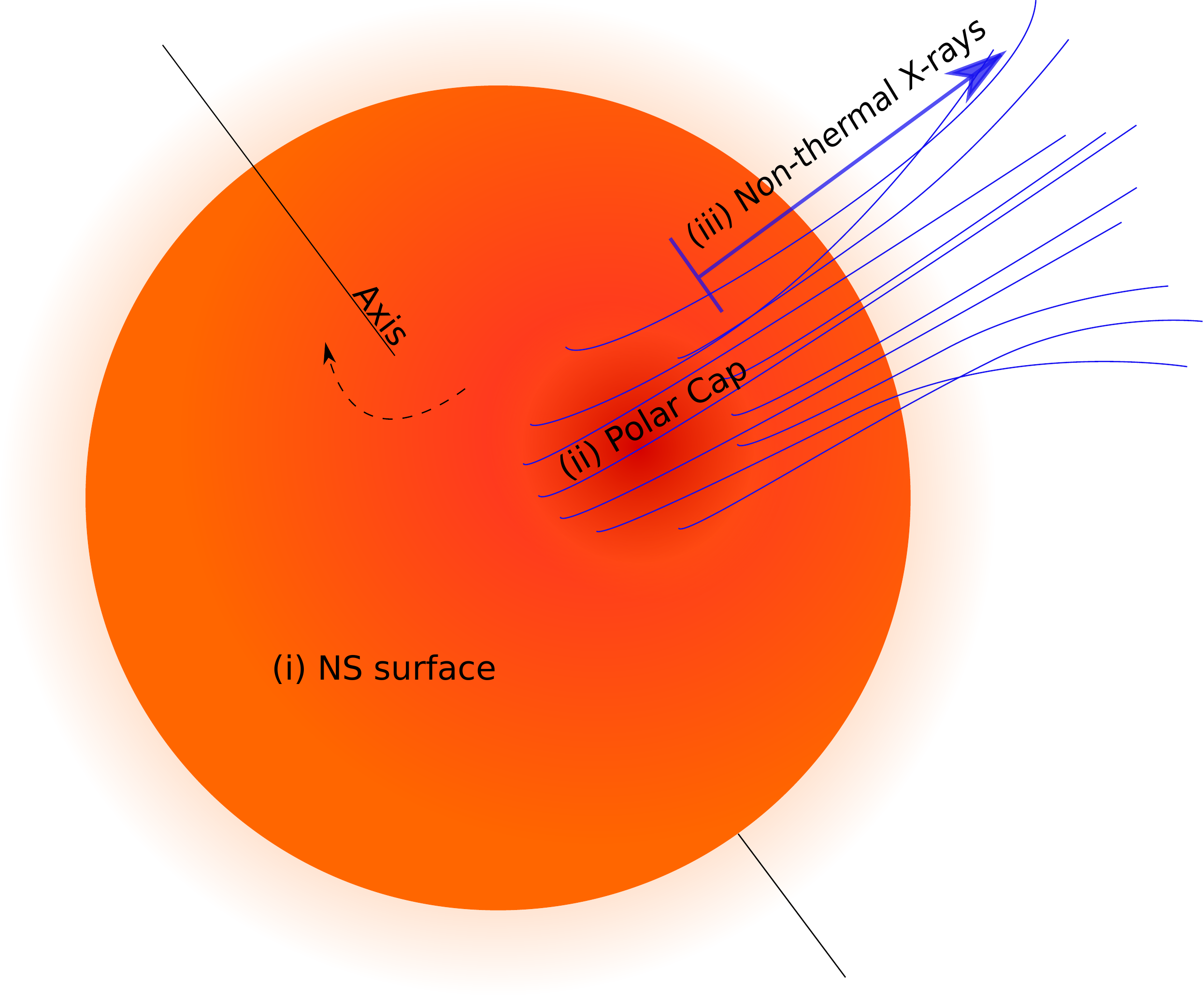}}
\subfloat[]{\label{fig:OldNSchematic}\includegraphics[width = 0.33\textwidth]{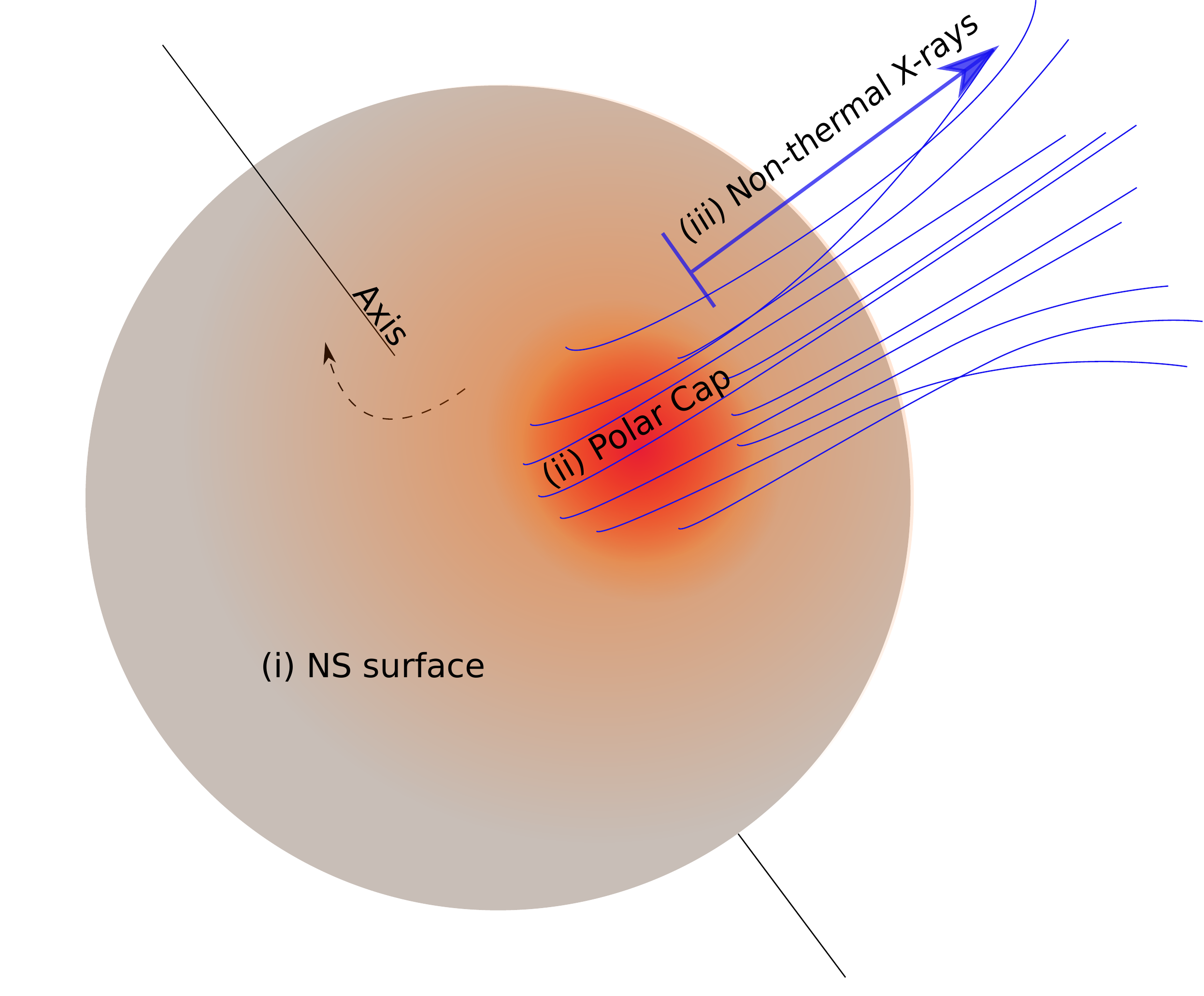}}
\renewcommand{\thesubfigure}{b-ii}
\subfloat[]{\label{fig:OldNSpec1}\includegraphics[width = 0.33\textwidth]{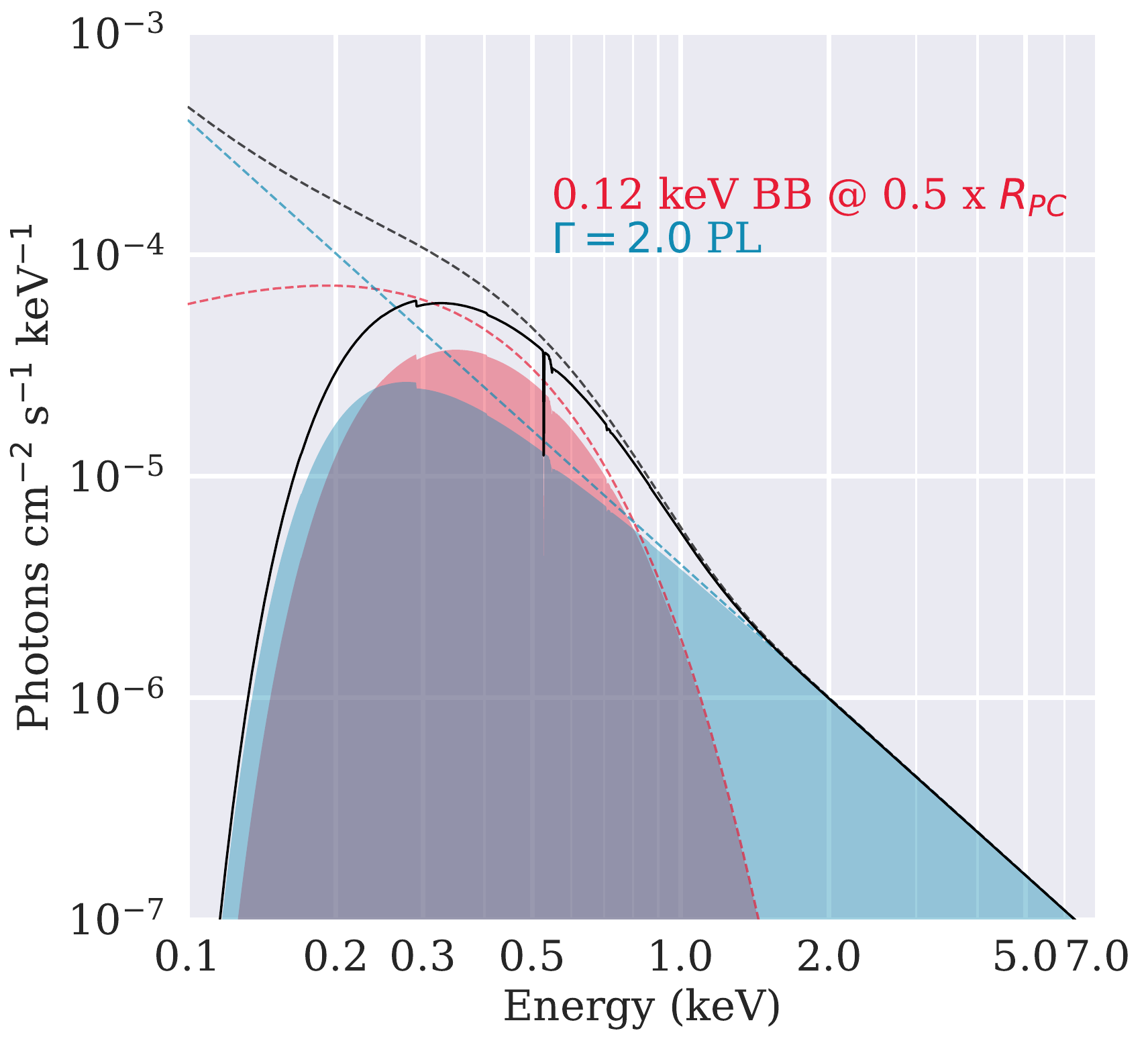}}\\
\renewcommand{\thesubfigure}{a-i}
\subfloat[]{\label{fig:NewNSpec}\includegraphics[width = 0.33\textwidth]{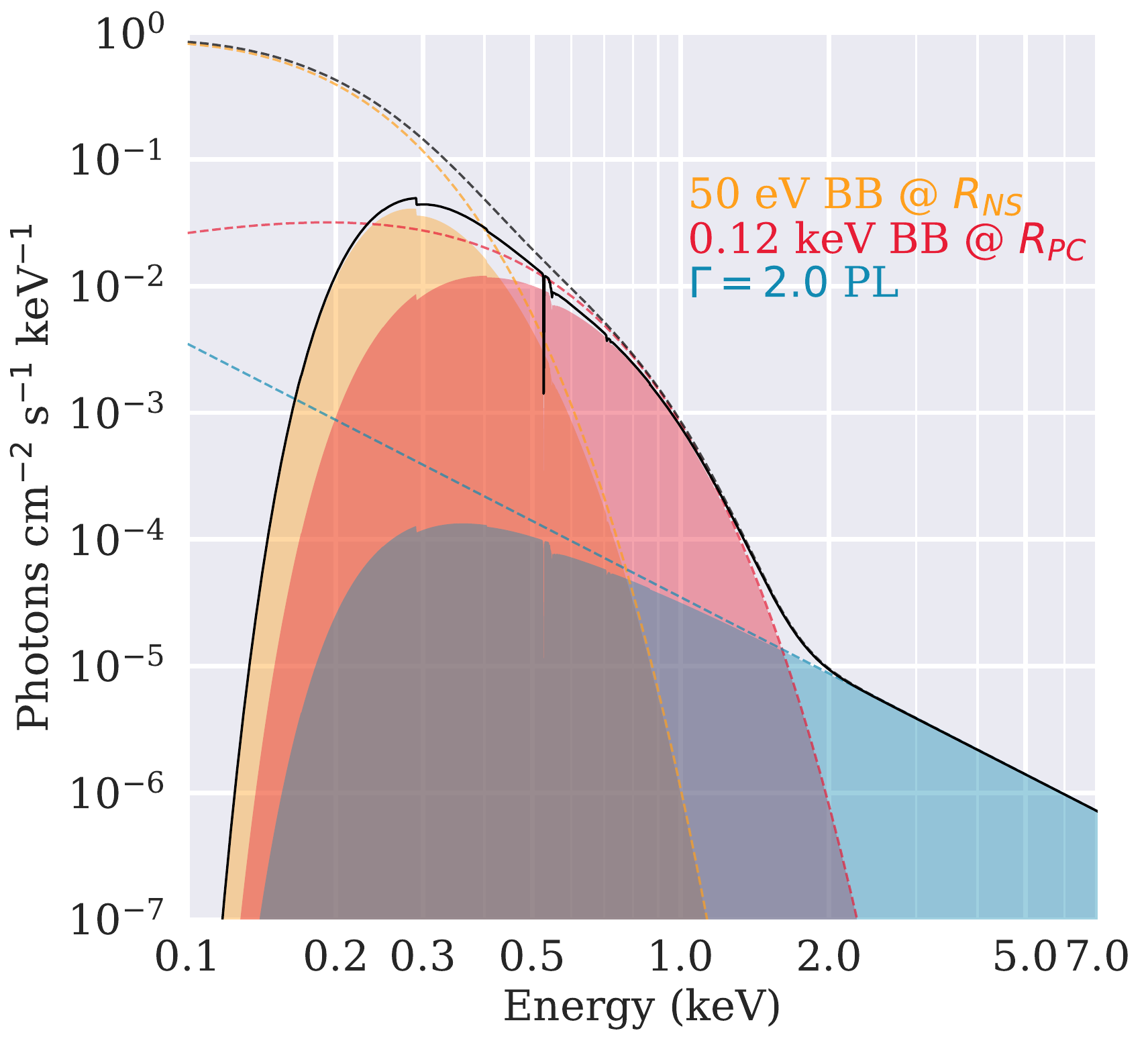}}
\renewcommand{\thesubfigure}{b-i}
\subfloat[]{\label{fig:OldNSpec}\includegraphics[width = 0.33\textwidth]{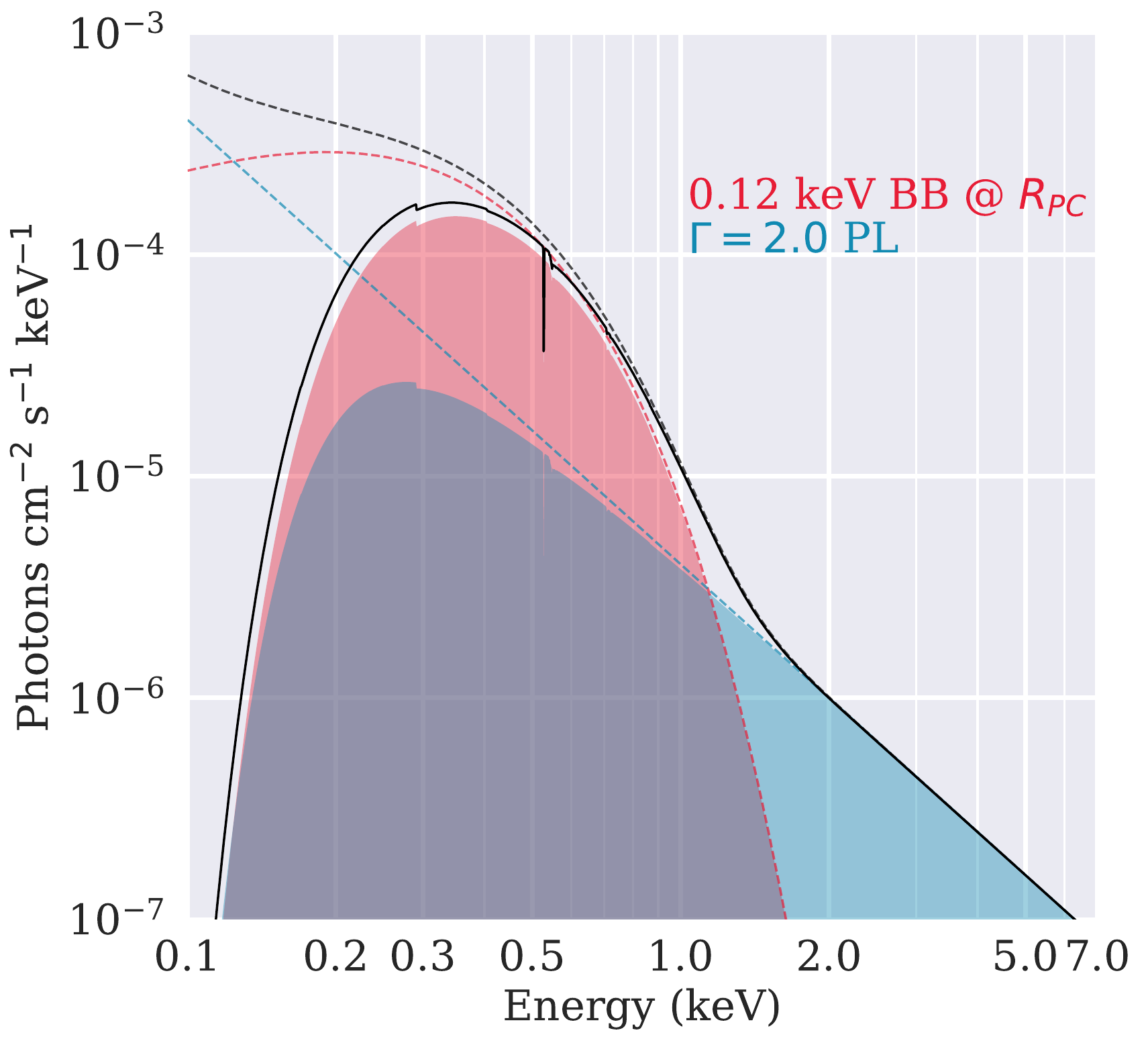}}
\renewcommand{\thesubfigure}{b-iii}
\subfloat[]{\label{fig:OldNSpec2}\includegraphics[width = 0.33\textwidth]{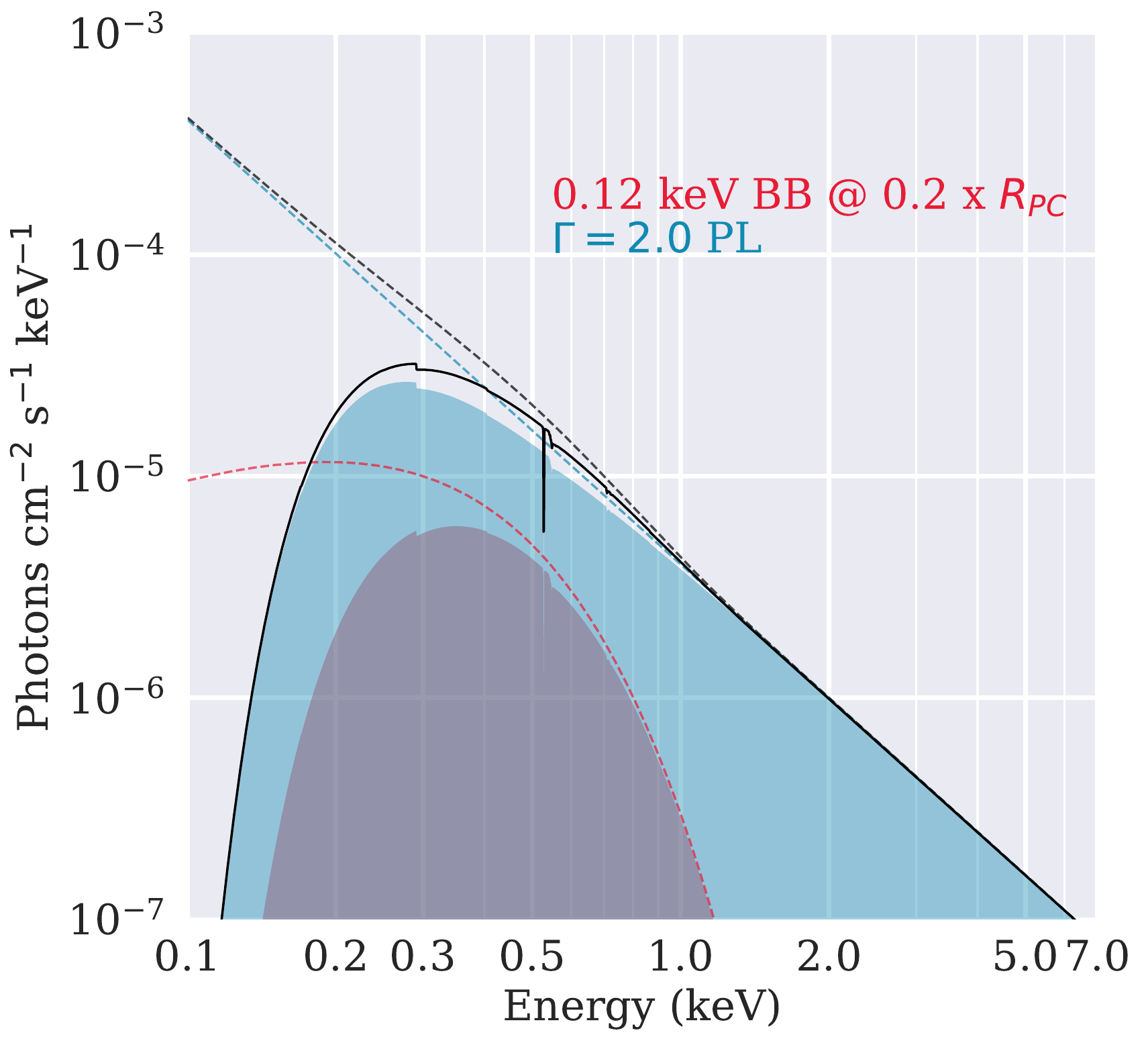}}
\caption{Schematic diagram (not to scale) illustrating the sites of X-ray emission, the bulk surface, the polar cap, and the open-field lines, in a young (a) and an old pulsar (b).
The region within the open-field lines where the non-thermal emission originates is not constrained.
Sample spectral models from a typical young pulsar (a-i), and an old pulsar with the thermal component from regions of different effective radii, $\sim$ 100\% (b-i), 50\% (b-ii), and 20\% (b-iii) of the dipole PC radius are also shown.
The filled curves and dashed lines show the contributions from each component, thermal from NS surface (orange) and hot-spot (red) and non-thermal (blue), with and without ISM absorption, respectively.
The black line shows the integrated spectral model after absorption.
}
\label{NSchematic}
\end{figure*}

{}X-ray emission from neutron stars can originate from a few different sites:\\
{}(i) The heat of formation and any subsequent internal re-heating is lost via thermal radiation from the bulk of the stellar surface (Figure \ref{fig:NewNSchematic}).
{}NS thermal evolution models predict cooling times of the order of 1 Myr beyond which the surface temperature is too low to produce significant X-rays (\citealt{Gnedin2001,YAKOVLEV2005,Page2006}).
{}Thermal soft X-rays from regions of sizes comparable to the NS radius have been observed from young pulsars (e.g., Geminga; \citealt{Halpern1993}, PSR B1055--52; \citealt{Oegelman1993}, PSR B0656+14; \citealt{Possenti1996}, Vela; \citealt{Pavlov_2001}).\\
{}(ii) The polar cap region on the NS surface is heated by highly-relativistic charged particles that are accelerated towards the surface (Figures \ref{fig:NewNSchematic},\ref{fig:OldNSchematic}).
{}X-ray detected pulsars, both young and old, often show a relatively hotter and localized thermal component in their spectra (e.g., see \citealt{DeLuca2005}).
{}This apparent hot spot is often identified with the polar cap.
{}The polar cap heating continues as long as the NS can accelerate particles in the inner vacuum gap (\citealt{Ogelman1991}; \citealt{Cheng1980}).\\
{}(iii) Charged particles accelerated in regions/gaps inside the open field lines can emit X-rays via synchrotron \citealt{Zhang1998} or possibly inverse-Compton radiation \citep{Lyutikov2013}.
{}Alternatively, certain regions within the return current can also accelerate charged particles and produce X-ray emission  (\citealt{Cerutti2017}; and references therein).
{}Non-thermal emission is very commonly observed in the X-ray spectra of RPP (\citealt{LiLuLi2008}, and references therein).

{}A phase-integrated spectrum is obtained by using source events from all detected rotational phases.
{}The individual emission components are better resolved when each of them dominate at a different energy range.
{}Figure \ref{fig:NewNSpec} illustrates a case wherein, for a young pulsar, the spectrum is dominated by the thermal emission from the bulk surface at low energies ($\lesssim0.5$ keV), the PC hot spot emission in the intermediate energies ($\approx0.5-1.5$ keV), and the non-thermal emission at higher energies ($\gtrsim1.5$ keV).
{}For an old pulsar, an ideal scenario (Figure \ref{fig:OldNSpec}) will have the thermal hot-spot emission dominating the lower energies ($\lesssim1$ keV) and the non-thermal component dominating the higher energies ($\gtrsim1$ keV).
{}For the purposes of this illustration, we use BB to represent the thermal component and PL for the non-thermal.

{}The spin-down power of pulsars reduces by as much as $\sim3-4$ orders of magnitude over the $\sim10^7$ years decrease in characteristic age, as seen from the $P-\dot{P}$ diagram of pulsars.
{}Additionally, as the pulsar slows down, the light-cylinder radius increases resulting in reduced polar cap area.
{}As a result, the emission from old pulsars is considerably weaker than those of younger pulsars for comparable distances, assuming no increase in PC heating.
{}As explained earlier, observed thermal emission area can be considerably smaller than the PC area for a dipole due to the orientation/geometry of the pulsar and the presence of multipolar magnetic fields.
{}This could lead to scenarios where the thermal component in the spectra dominates only in a very narrow energy range (Figure \ref{fig:OldNSpec1}) or is simply too weak compared to the non-thermal component  (Figure \ref{fig:OldNSpec2}).
{}In such cases, with low S/N data as is typically the case, the phase-integrated spectrum cannot differentiate between a two-component thermal+non-thermal emission model and a single non-thermal emission model (e.g., J0108 and the set of old NSs listed above).

{}The source, J0108, is detected only in the $0.15-2$ keV range, suggesting a soft spectrum.
{}The EPIC-pn and MOS spectral extraction parameters optimized over GTIs and extraction aperture are shown in Table \ref{tab:extraction}.
{}We group the spectra with at least 1 count per energy bin while ensuring that the bin width is at least 1/3 of the detector resolution full width at half maximum (FWHM).

{}We perform spectral fitting using XSPEC 12.10.0c.
{}We model the spectrum with a simple power-law (PL; \texttt{powerlaw} in XSPEC), modeling the interstellar absorption using the T\"{u}bingen-Boulder model (\texttt{Tbabs} in XSPEC; \citealt{Wilms2000}) with solar abundance table \texttt{wilm} from \cite{Wilms2000}, and photoelectric cross-section table \texttt{bcmc} (\citealt{Balucinska1992,Yan1998}).

\begin{figure}
\captionsetup[subfigure]{labelformat=empty}
\subfloat[]{\label{fig:J0108Spb}\includegraphics[width = 0.47\textwidth]{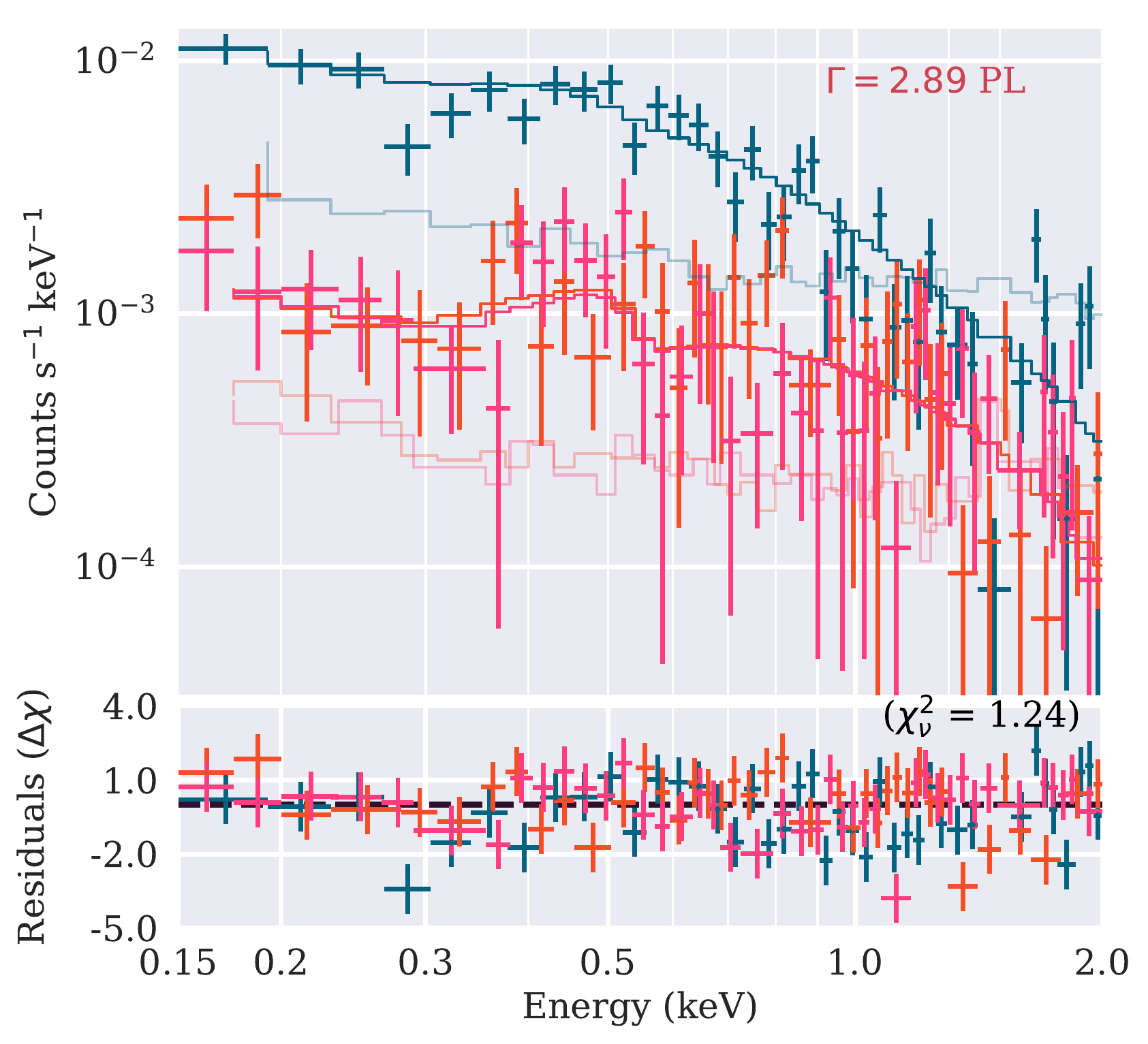}}
\caption{PL model fit to the phase-integrated source spectra.
Shown are the spectra (error-bars), best-fit model (steps), and the residuals (bottom panel) for EPIC-pn (blue), MOS1 (pink), and MOS2 (orange) spectra.
The observed background spectra are overlaid as steps (without error-bars for clarity) using the same color codes.
The annotated photon index $\Gamma$ is the best-fit value, which is slightly different from the median of the marginal distribution.}
\label{J0108ISpectrum}
\end{figure}

\begin{figure}
\centering
\includegraphics[width = 0.47\textwidth]{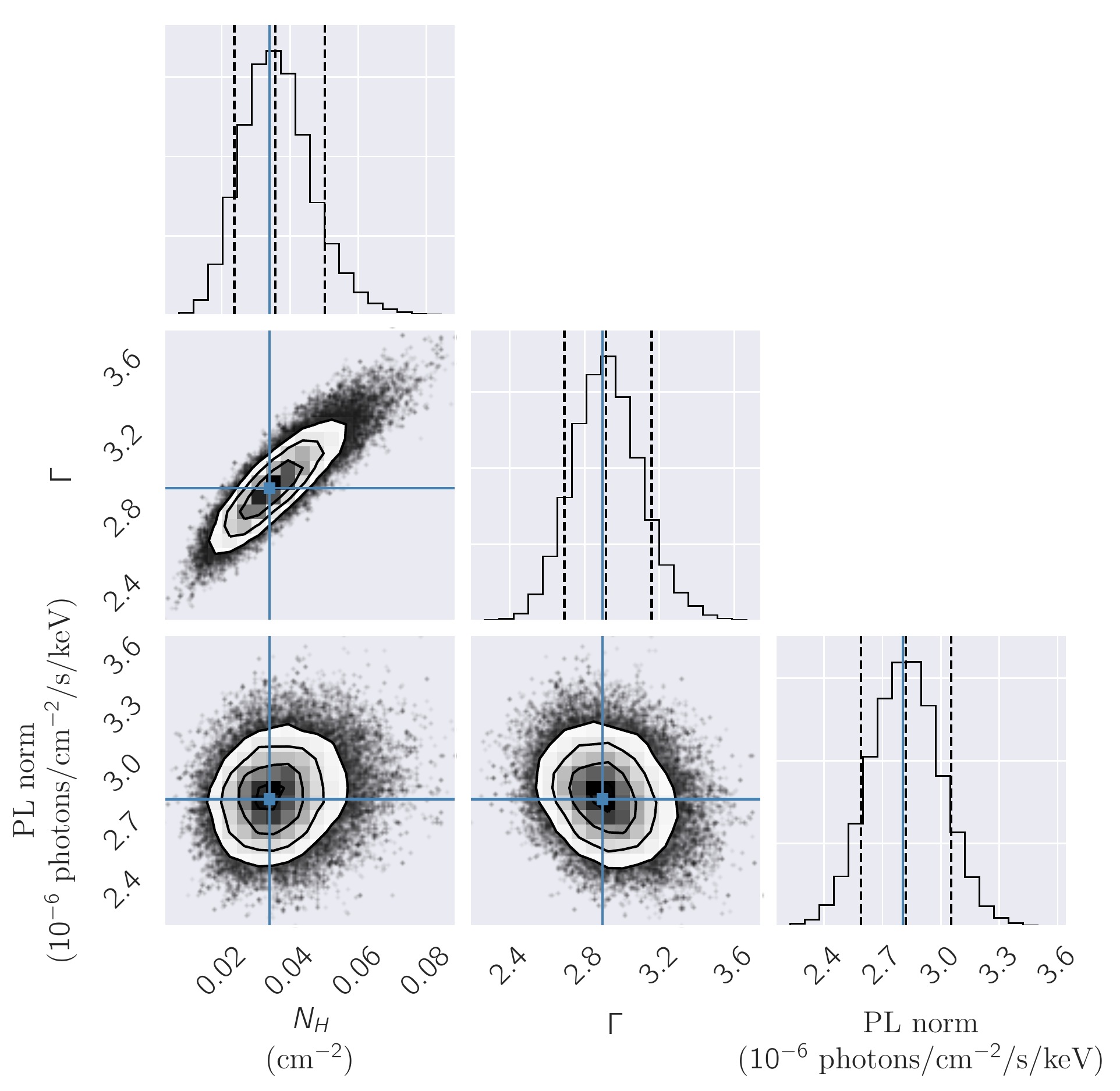}
\caption{\label{fig:PhIcorner} 1D marginalized distributions and 2D marginalized joint-plots between the PL fit parameters to the phase-integrated spectrum.
The parameters corresponding to the highest posterior probability are shown with blue lines.
The dashed lines in the 1D distributions show the 16, 50, and 84 percentile limits.
The contour lines in the 2D joint plots show the $1,2,\text{and}\,3-\sigma$ bi-variate uncertainties for pairs of parameters.}
\end{figure}

{}We perform Bayesian fits to the spectra using the statistic \texttt{lstat}\footnote{\url{https://heasarc.gsfc.nasa.gov/xanadu/xspec/manual/XSappendixStatistics.html}} in XSPEC which constructs the likelihood function by combining the source and background counts (without subtraction) assuming Poisson distribution for both.
{}The posterior probability distribution is marginalized over the background rate parameter to retain only the distributions of the emission model parameters.
{}We select the Goodman-Weare algorithm \citep{Goodman2010} and set up Monte Carlo Markov chain (MCMC) with 50 walkers taking 2000 steps each (1000 steps for burn-in) to sample the posterior probability distribution over the model parameter space.
{}The PL fit is similar to the one obtained in BP2012.
{}A $\Gamma=3$ PL fits the spectra well with no significant systematics in the residuals.
{}In Figure \ref{J0108ISpectrum}, we show the spectral fit with residuals and in Figure \ref{fig:PhIcorner}, we show the marginalized individual and joint posterior probability distributions of the fit parameters.

{}Other single-component models such as blackbody (BB: \texttt{bbodyrad} in XSPEC) and neutron star atmosphere (NSA: \texttt{nsa} or NSMAXG: \texttt{nsmaxg} models in XSPEC) fit worse than PL, with high residuals in the high energy bins ($\gtrsim 1$ keV).
{}Additional components to the PL are not required to explain the phase-integrated spectrum, due to the low S/N of the data.
However, an additional thermal component (BB) is allowed with a very low probability and seen to dominate the emission at energies $\lesssim 0.8$ keV.
We conclude that the phase-integrated spectra of J0108, due to their low S/N and possibly a relatively weaker thermal component, cannot confirm or reject the canonical model of two-component emission from old pulsars.

\begin{table*}
\centering
\ra{1.1}
\caption{EPIC spectral extraction parameters for PSR J0108--1431.}
\label{tab:extraction}
\begin{threeparttable}
\begin{tabular}{@{}lccccc@{}}\toprule
Detector: & \multicolumn{3}{c}{EPIC-pn} & MOSI &  MOSII\\
\cmidrule{2-4}
Phase range $\phi:$ & $0-1$ & $0.2-0.7$ & $0.7-0.2$ & $0-1$ & $0-1$\\ \midrule
Energy range (keV) & $0.15-2$ & $0.15-1.2$\tnote{a} & $0.15-2$ & $0.15-2$ & $0.15-2$ \\
Net exposure (ks) & 110 & 55 & 55 & 126 & 126 \\
Radius & $13\arcsec$ & $12\arcsec$ & $13\arcsec$ & $12\arcsec$ & $12\arcsec$ \\
Source counts & 922 & 309 & 579 & 218 & 197 \\
Net count rate (ks$^{-1}$) & $5.50\pm0.28$ & $3.87\pm0.32$ & $6.74\pm0.44$ & $1.25\pm0.12$ & $1.13\pm0.11$ \\
S/N ratio & 20 & 12 & 15 & 11 & 10 \\
\bottomrule
\end{tabular}
  \begin{tablenotes}
    \item[a] The pulsar has no contribution above 1.2 keV in this phase range.
  \end{tablenotes}
\end{threeparttable}
\end{table*}

\begin{table*}
\centering
\caption{Marginalized parameter distributions represented by medians with $10-90$ percentile intervals from PSR J0108--1431 spectral fits.}
\label{tab:spectralfit}
\begin{threeparttable}
\begin{tabular}{@{}lc|>{\columncolor[gray]{0.95}}c|cc|>{\columncolor[gray]{0.95}}c@{}}\toprule
& \multicolumn{5}{c}{Phase-range} \\
\cmidrule{2-6}
Parameter & $0-1$ & $0.2-0.7$ & $0.2-0.7$ & $0.7-0.2$ & $0.7-0.2$\\
\midrule
$N_{H}$ ($10^{21}$ cm$^{-2}$) & $0.36^{+0.15}_{-0.12}$  & $1.46^{+1.15}_{-0.64}$  & \multicolumn{2}{c|}{$0.21^{+0.17}_{-0.12}$}  & $0.022_{-0.019}^{+0.052}$ \\[1.0EX]
$\Gamma$ & $2.93^{+0.25}_{-0.23}$  & $5.0^{+1.4}_{-1.0}$  & ---  & $2.45^{+0.33}_{-0.27}$  & ---  \\[1.0EX]
PL norm ($N_{-6}$)\tnote{a} & $1.62^{+0.51}_{-0.42}$  & $1.51^{+0.83}_{-0.60}$  & ---  & $3.55^{+0.44}_{-0.42}$  & ---  \\[1.0EX]
$F^{\rm unabs.}_{0.3-7 {\rm keV}}$\tnote{b} & $1.41^{+0.15}_{-0.12}$  & $3.1^{+8.1}_{-1.6}$  & ---  & $1.67^{+0.24}_{-0.20}$  & ---  \\[1.0EX]
$kT_{BB}$ (eV) & ---  & ---  & $115^{+12}_{-12}$  & ---  & $164_{-13}^{+14}$ \\[1.0EX]
BB norm\tnote{c} & ---  & ---  & $ 5.6^{+4.9}_{-2.3}$  & ---  & $1.48_{-0.41}^{+0.60}$ \\[1.0EX]
BB area ratio\tnote{d} & ---  & ---  & $ 0.38^{+0.34}_{-0.16}$  & ---  & $0.10_{-0.03}^{+0.04}$ \\[1.0EX]
$L_{BB, bol}$ ($10^{28}$ erg s$^{-1}$) & ---   & ---   & $5.62^{+1.59}_{-0.98}$   & ---  & $6.16_{-0.52}^{+0.57}$ \\[1.0EX]
$\chi_\nu^2$ & 1.2 & 1.5 &  \multicolumn{2}{c|}{1.2\tnote{e}} & 1.7 \\
\bottomrule
\end{tabular}
  \begin{tablenotes}
    \item The fits corresponding to the parameters in the grey columns are statistically unacceptable.\\
    \item[a] Power-law normalization in units of $10^{-6}$ photons cm$^{-2}$ s$^{-1}$ keV$^{-1}$ at 1 keV.\\
    \item[b] Unabsorbed PL flux in the $0.3-7$ keV range.\\
    \item[c] Blackbody normalization $=R^2_{\rm BB}/d^2_{\rm 10 kpc}$, where $R_{\rm BB}$ is the effective radius of the BB emission region in km, and $d_{\rm 10 kpc}$ (= 0.021 for J0108) is the distance in units of 10 kpc.\\
    \item[d] Ratio of observed BB area to the theoretical polar cap area for a dipolar magnetic field ($A^\infty_{\rm d,PC}=0.139$ km$^2$).\\
    \item[e] Spectra in the phase ranges $0.2-0.7$ and $0.7-0.2$ are simultaneously fit with BB and PL models, respectively, while using a common ISM absorption parameter $N_H$.
  \end{tablenotes}
\end{threeparttable}
\end{table*}

\section{X-ray pulsations from the pulsar}\label{sec:pulsations}

{}The 73.4 ms time-resolution of the EPIC-pn detector allows us to assign rotational phase of photon events with an uncertainty $\Delta\phi\approx0.09$.
Following the events extraction procedure explained in section \ref{sec:dataAnalysis}, we extracted 1057 counts from the source region in the $0.15-2$ keV energy range with different GTIs and extraction radii for each energy bin (Table \ref{tab:timingex}).
This ensures high S/N events extraction in the selected energy bins.
The events were barycentered using the SAS task \texttt{barycen}.
We use the $Z_n^2$ test (for up to n=4 harmonics) to find the pulsation frequency within a 0.002 Hz window around the pulsar frequency $\nu_{\rm exp} = 1.2382909579(8)$ Hz, extrapolated from the known ephemeris at MJD 50889, where, the $1\sigma$ uncertainty in the last digit is given in parenthesis.
We detect pulsations at $\nu = 1.23829043(53)$ with $Z_2^2 = 30.2$ ($4.4\sigma$) and statistically significant contributions up to 2 harmonics.
The detected frequency is within $1.5\sigma$ of the expected frequency.
For the observation span of 127 ks, $Z_n^2$ is only sensitive to frequency derivative $\gtrsim 8\times10^{-6}$ Hz s$^{-1}$ and frequency second derivative $\gtrsim 8\times10^{-16}$ Hz s$^{-2}$, both of which are orders of magnitude greater than the pulsar's frequency derivative and second derivative (see Table \ref{tbl-1_summary}).

\begin{table}
\centering
\caption{Timing extraction parameters.}
\label{tab:timingex}
\begin{threeparttable}
\begin{tabular}{@{}lccccc@{}}\toprule
Energy Range & GTI & Radius & (Net/Total) & S/N & $Z_2^2$ \\
(eV) & (ks) & & & & \\
\midrule
$150-300$ & 127 & $22\farcs5$ & 244\;/\;413 & 11.6 & \\
$300-400$ & 127 & $22\farcs5$ & 122\;/\;190 & 8.6 & \\
$400-500$ & 127 & $10\farcs5$ & 81\;/\;93 & 8.3 & \\
$500-600$ & 127 & $10\farcs5$ & 61\;/\;72 & 7.2 & \\
$600-700$ & 118 & $15\farcs5$ & 54\;/\;67 & 6.6 & \\
$700-900$ & 127 & $12\farcs5$ & 67\;/\;95 & 6.8 & \\
$900-2000$ & 127 & $7\farcs5$ & 76\;/\;126 & 6.8 & \\\hline
$150-300$ & 92 & $7\farcs5$ & 83\;/\;95 & 8.5 & 28.8 \\
$300-500$ & 127 & $10\arcsec$ & 146\;/\;172 & 11.1 & 17.6 \\
$500-600$ & 84 & $5\arcsec$ & 22\;/\;23 & 4.6 & 24.9 \\
$600-1000$ & 109 & $22\farcs5$ & 131\;/\;226 & 8.5 & 12.8 \\
$1000-2000$ & 127 & $20\arcsec$ & 51\;/\;144 & 4.1 & 36.2 \\
\bottomrule
\end{tabular}
  \begin{tablenotes}
    \item The set of extraction parameters on the top maximize the S/N of the events whereas the set at the bottom maximizes the pulsations statistics ($Z_2^2$). The overall $Z_2^2$ for the extracted events is 30.2 when S/N is maximized and 71.7 when $Z_2^2$ in maximized.
  \end{tablenotes}
\end{threeparttable}
\end{table}

\begin{figure}
\captionsetup[subfigure]{labelformat=empty}
\subfloat[]{\label{fig:J0108Ta}\includegraphics[width=0.47\textwidth]{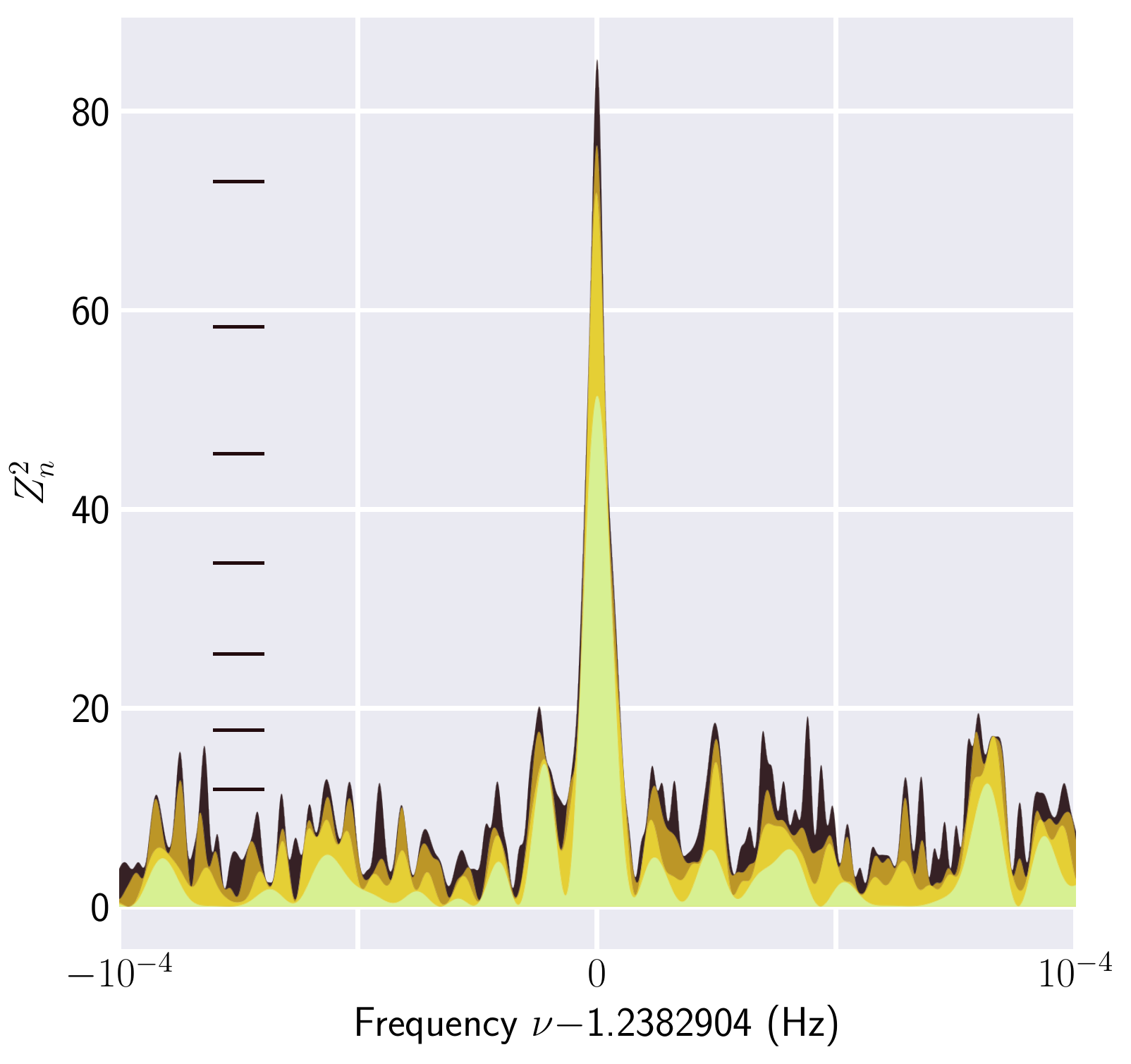}}\\[-4ex]
\subfloat[]{\label{fig:J0108Tb}\includegraphics[width=0.48\textwidth]{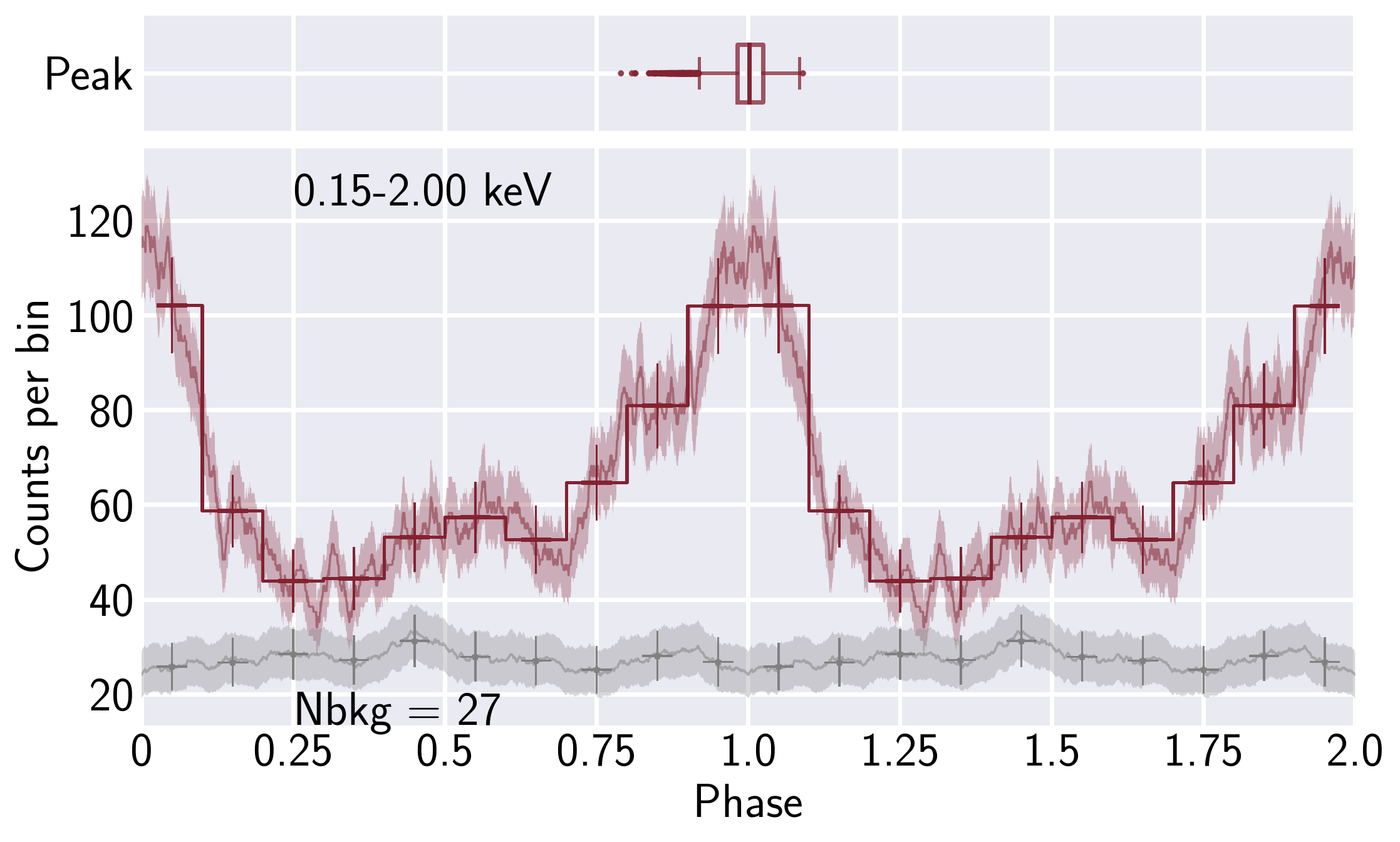}}
\caption{Top: $Z_{1-4}^2$ (light to dark shades) statistics obtained for $0.15-2$\,keV EPIC-pn events extracted from an energy-dependent, $Z_2^2$-maximizing source events extraction.
The dashes on the left of the plot mark the $1\sigma-7\sigma$ significance levels for the $Z_4^2$ test.
Bottom: Phase-folded $0.15-2$\,keV pulse profile shown as binned events histogram and a KDE smoothed profile using the events' frame-times as uniform density kernels.
The estimate of the background contribution is shown in grey and the annotation gives the average background contribution (Nbkg).
The bootstrap estimate of the peak phase is represented above each profile plot using box plots that show the three quartiles (box) and the probable limits of the distribution using $1.5\times$ inter-quartile range (whiskers).
Points beyond the whiskers are likely outliers and hence not part of the distribution.
}
\label{J0108Timing}
\end{figure}

In order to increase the pulsations significance, we repeated the source events extraction over the GTIs, energy ranges, and extraction aperture sizes by maximizing the $Z_2^2$ statistics instead of the conventional S/N maximization.
We extracted 660 counts from the source region in the $0.15-2$ keV range with varying GTIs for different energy bins, and extraction aperture radii varying between $5\arcsec-15\arcsec$.
These events show significantly stronger pulsations at $\nu = 1.23829044(25)$ with $Z_4^2 = 85.2$ ($7.8\sigma$) and statistically significant contributions up to 4 harmonics (Figure \ref{J0108Timing}).
The $Z_2^2 = 71.7$ is also significantly higher than the value obtained from simple S/N optimization.

Phase-folding the source and background events with the frequency of the detected pulsations, we obtain a binned pulse profile choosing MJD 55727.93394840 (XMM reference MJD 50814.0 + mission elapsed time 424563893.142/86400 days) as phase zero (Figure \ref{J0108Timing}).
The frame-time of the detector mode spans $\approx 0.09$ in phase and hence, for a profile with 10 bins, the uncertainty in phase for any event is close to the bin width.
To obtain a reliable binning-independent pulse profile, we use Kernel Density Estimation (KDE; \citealt{FeigelsonBabu2012}) smoothing using rectangular (boxcar function) kernels for each event with the edges matching the event frame-times.
This assigns a uniform probability of an event detection within its frame-time.
Such probability distributions for all events are then superimposed to obtain the probability density function (PDF) of event detection at any phase.
This PDF is multiplied with the average counts per bin of any binning scheme to obtain the corresponding smoothed profile.

\begin{figure}
\centering
\includegraphics[width=0.47\textwidth]{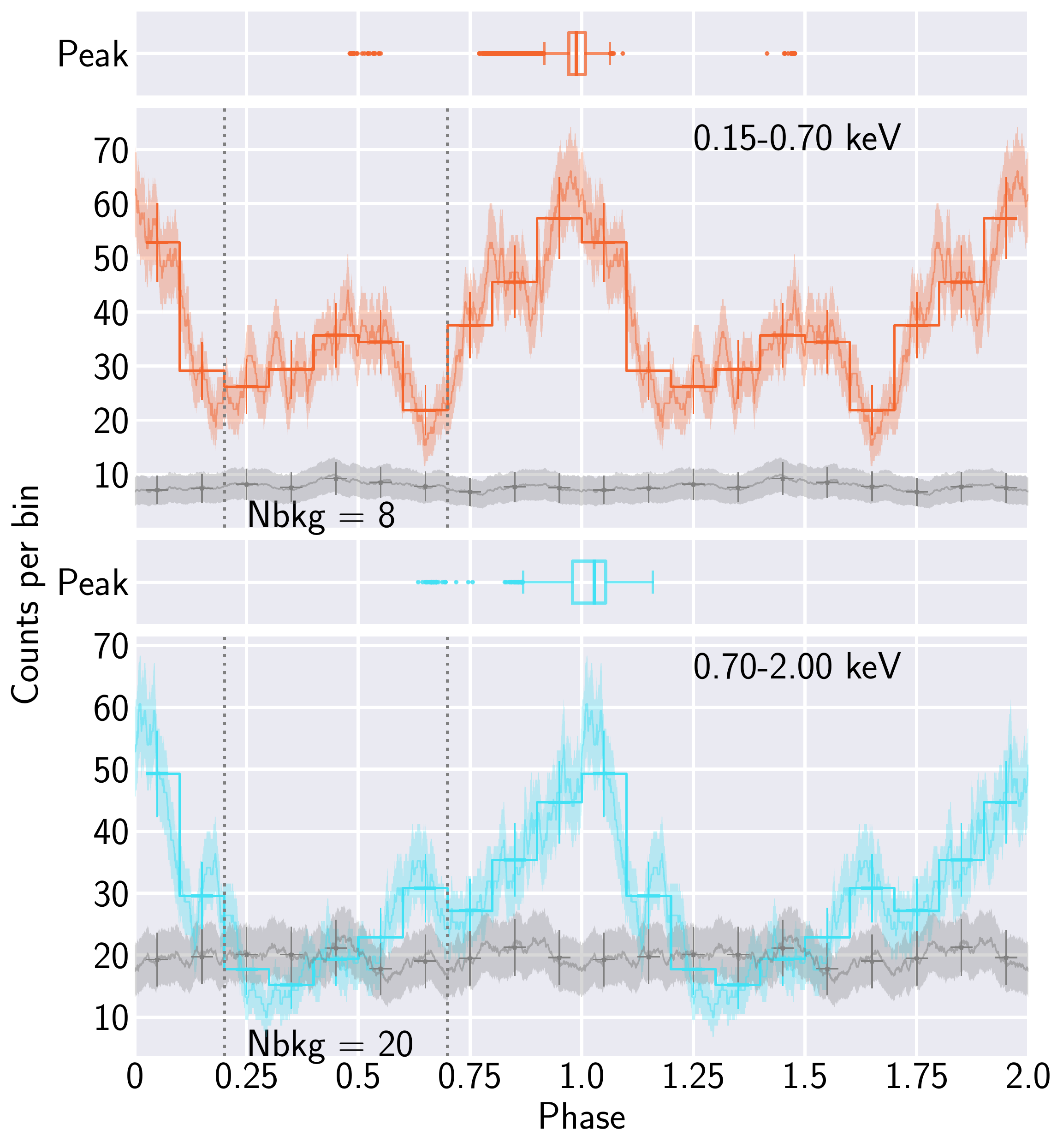}
\caption{\label{fig:eVtiming}Top to bottom: Pulse profiles in the $0.15-0.7$\,keV and $0.7-2$\,keV energy ranges.
The KDE smoothing is performed using the events' frame-times as uniform density kernels.
The estimate of the background contributions is shown in grey.
The box plots for the peak phase follows the same convention as Figure \ref{J0108Timing}.
The vertical dotted lines separates the two phase ranges $0.2-0.7$ and $0.7-0.2$ used in phase-separated spectral analysis.
}
\end{figure}

The binned and smoothed phase-integrated profile in the 0.15-2 keV range is shown in Figure \ref{J0108Timing}.
The profile is strongly peaked around phase $\phi=0$ for our chosen folding time reference.
Such narrow-peaked (non-sinusoidal) profiles are characteristic of non-thermal emission or highly beamed thermal emission.
BB emission, on the other hand, shows broad sine-like pulse with large duty cycle when special and general relativistic effects are included.
We attempted to detect thermal pulsations assuming the presence of a thermal component in the phase-integrated spectra.
In trial spectral fitting with combined PL+BB model, the thermal component dominates at energies $\lesssim0.8$ keV.
Hence, we produced energy-resolved pulse profiles in the $0.15-0.7$ keV, and $0.7-2$ keV ranges (Figure \ref{fig:eVtiming}).

We estimate the phase of pulse profiles and its uncertainties using Bootstrap resampling \citep{FeigelsonBabu2012}, assuming Poisson distributed counts in each bin and taking the observed values to be the mean.
The peak phase estimate is shown as Box plots in Figures \ref{fig:J0108Tb} and \ref{fig:eVtiming}), where the bars on the box represent the three quartiles (25\%, 50\%, and 74\%), the whiskers connected to the box at $1.5\times$ the inter-quartile range represent the probable limits of the distribution and the points beyond are the outliers.
The overall pulse peak is maintained at $\phi\approx0$ in the selected energy ranges with no statistically significant shift in peak phase.

In addition to the main pulse seen at all energy ranges near phase 0, the profiles show a possible second pulse around phase 0.5 which is dominant only in the $0.15-0.7$ keV range.
This hints at the possibility of a separate soft-emission component.
We assess the significance of the overall profile change in the two energy ranges by applying the Anderson-Darling test \citep{Anderson1954} to the binned profiles.
The null hypothesis that profiles are samples obtained from the same underlying distribution is not rejected, possibly due to the low number of counts.
Hence, the change in profile from double-peaked in $0.15-0.7$ keV to single-peaked between $0.7-2$ keV is not sufficient to infer two phase-separated emission components.
However, we use spectra extracted from the two phase ranges, $0.2-0.7$ and $0.7-0.2$ to further explore the emission components.

\section{Phase-separated Spectral Components}

From the pulse profile shapes, we inferred soft emission in the $0.2-0.7$ phase range and relatively harder emission in the rest of the phases ($0.7-0.2$).
We extract spectra in these two broad phase ranges from the EPIC-pn detector.
A finer phase-resolved spectral extraction is not reliable due to the low number of source counts and the coarse time-resolution.
The spectrum extracted from the $0.2-0.7$ phase range (S1) has $213\pm18$ net source counts in the $0.15-1.2$ keV range and none over 1.2 keV (Table \ref{tab:extraction}).
The spectrum extracted from the $0.7-0.2$ phase range (S2), on the other hand, has $371\pm24$ net source counts in the $0.15-2$ keV range.

\subsection{Blackbody Emission Model}\label{subsec:BBspectral}
\begin{figure*}
    \begin{minipage}[b]{.47\textwidth}
        \centering
        \includegraphics[width=\textwidth]{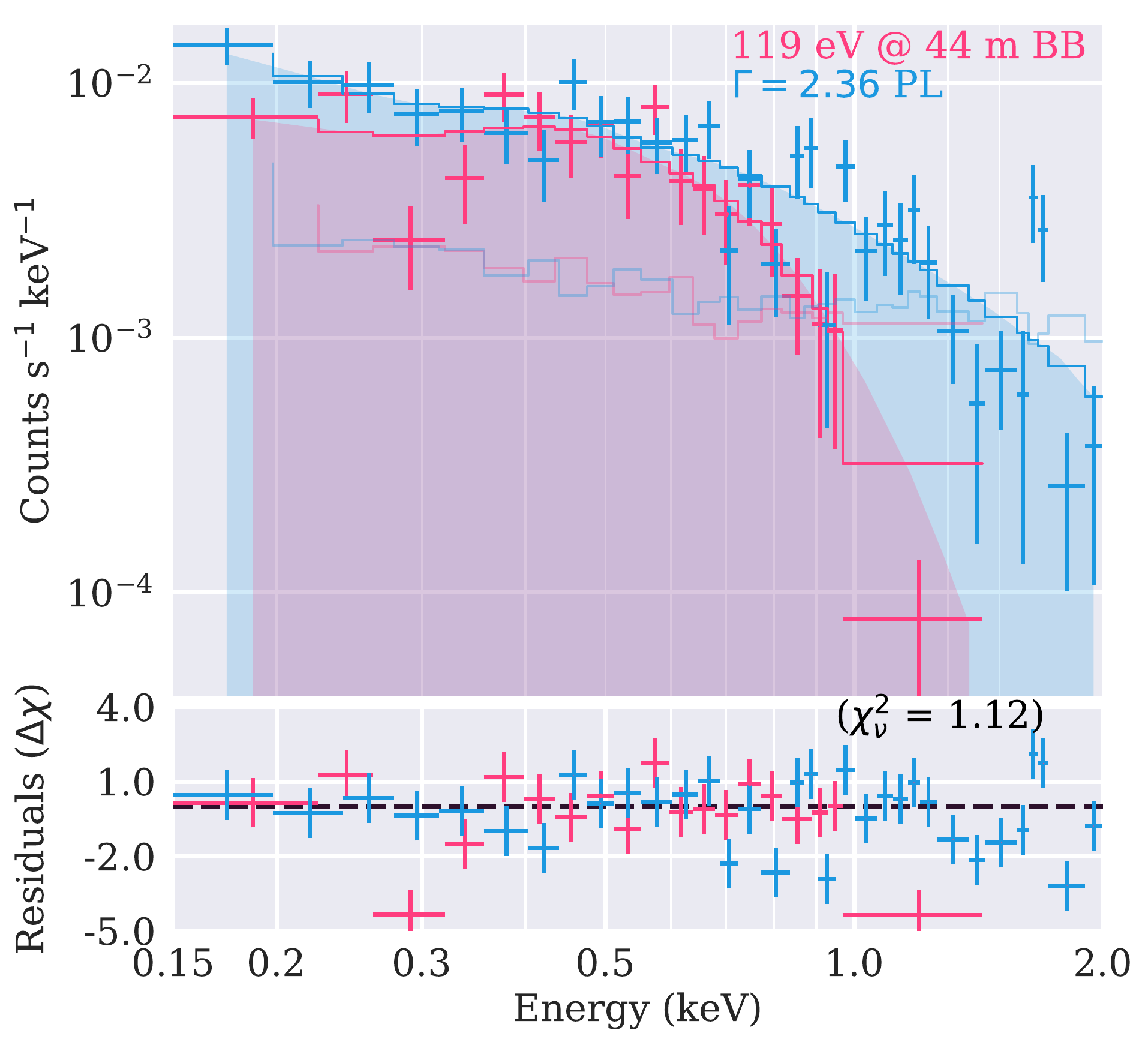}\\
        \includegraphics[width=\textwidth]{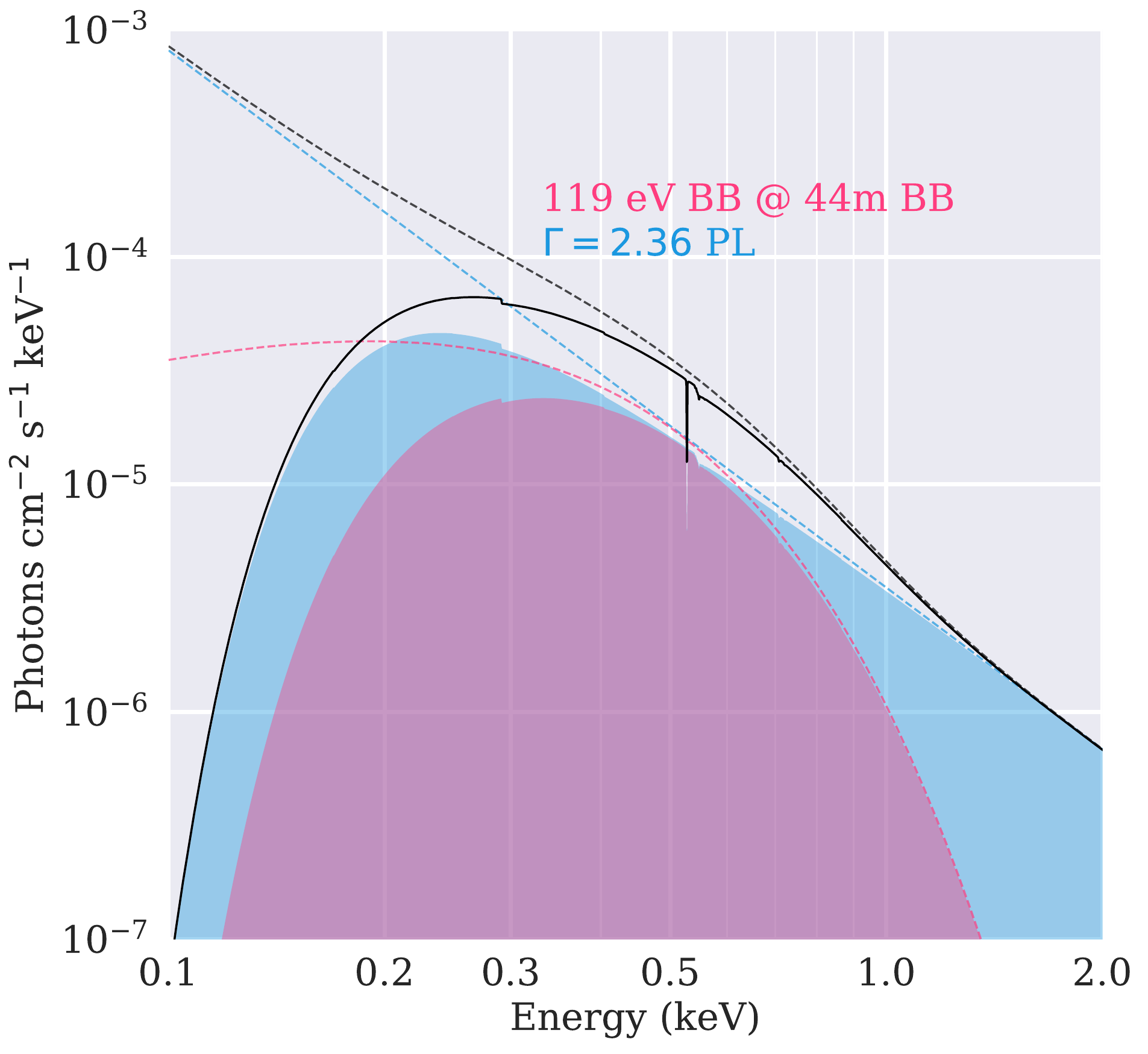}
    \end{minipage}
    \hfill
    \begin{minipage}[b]{.48\textwidth}
        \centering
        \includegraphics[trim=0 0 5.5cm 0,clip,width=\textwidth]{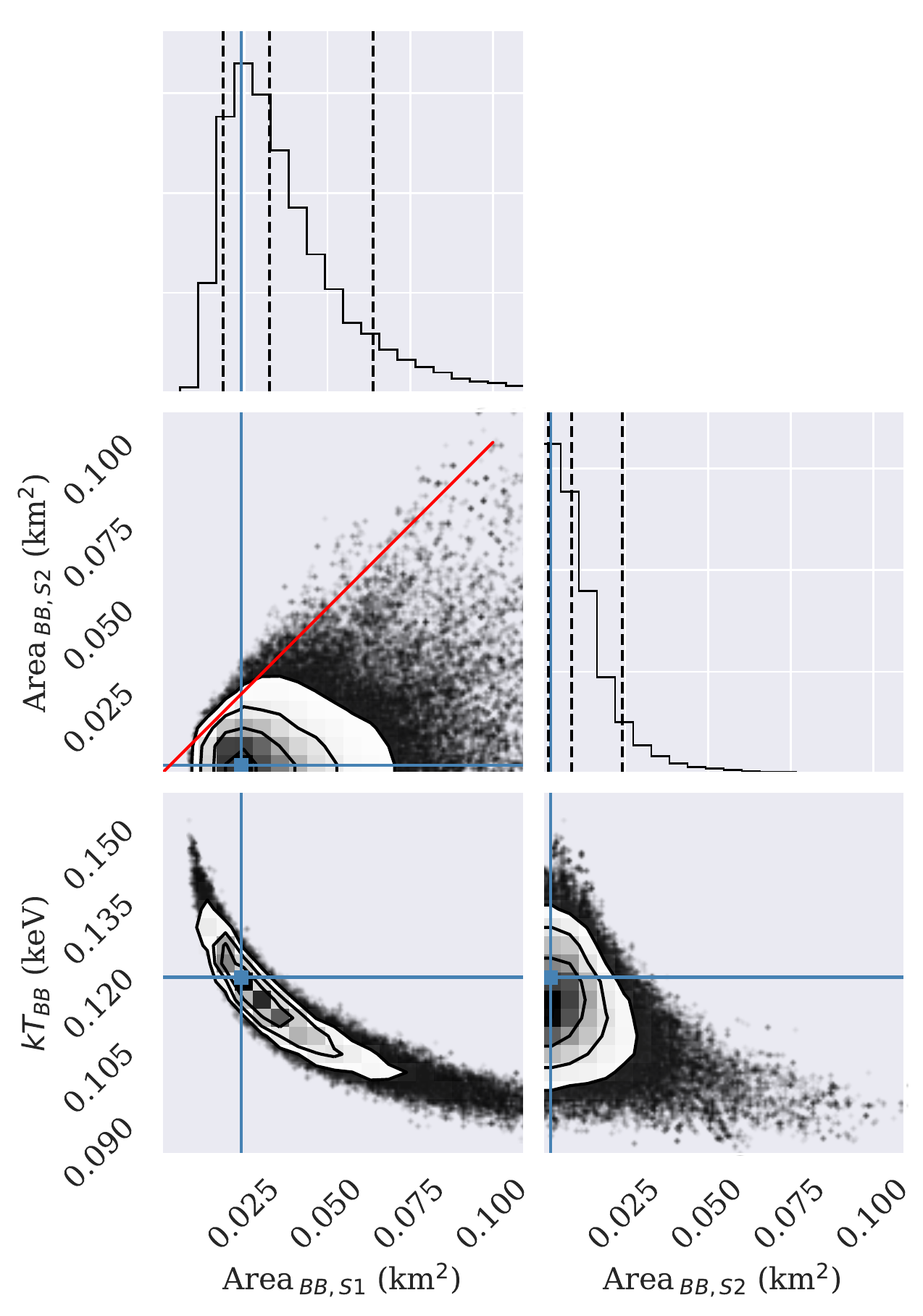}
    \end{minipage}  
\caption{Top-Left: Simultaneous fit to spectra from phase ranges $0.2-0.7$ (red) and $0.7-0.2$ (blue) using BB for the former and PL for the latter while using a tied $N_H$ parameter.
The best fit models (values annotated) are shown with bold steps over the data and the background contribution with thin steps.
Residuals are shown in the middle panel.
Bottom-Left: The ideal model spectral components (PL in blue, BB in red, and the combination in yellow) are shown with (filled or bold curves) and without (dashed lines) including the interstellar medium absorption model.
Right: Posterior pair plots between PC BB temperature ($kT_{BB}$) and effective BB area observed in the phase ranges $0.2-0.7$ (Area$_{BB,S1}$) and $0.7-0.2$ (Area$_{BB,S2}$ Area).
The 10-50-90 percentile contours/limits are shown for the two-dimensional distributions and single parameter marginal plots.
The blue lines correspond to parameters with the highest posterior probability and the red diagonal is a line of slope = 1.}
\label{fig:J0108PhRSpectrum}
\end{figure*}

\begin{figure*}
\centering
\includegraphics[width = \textwidth]{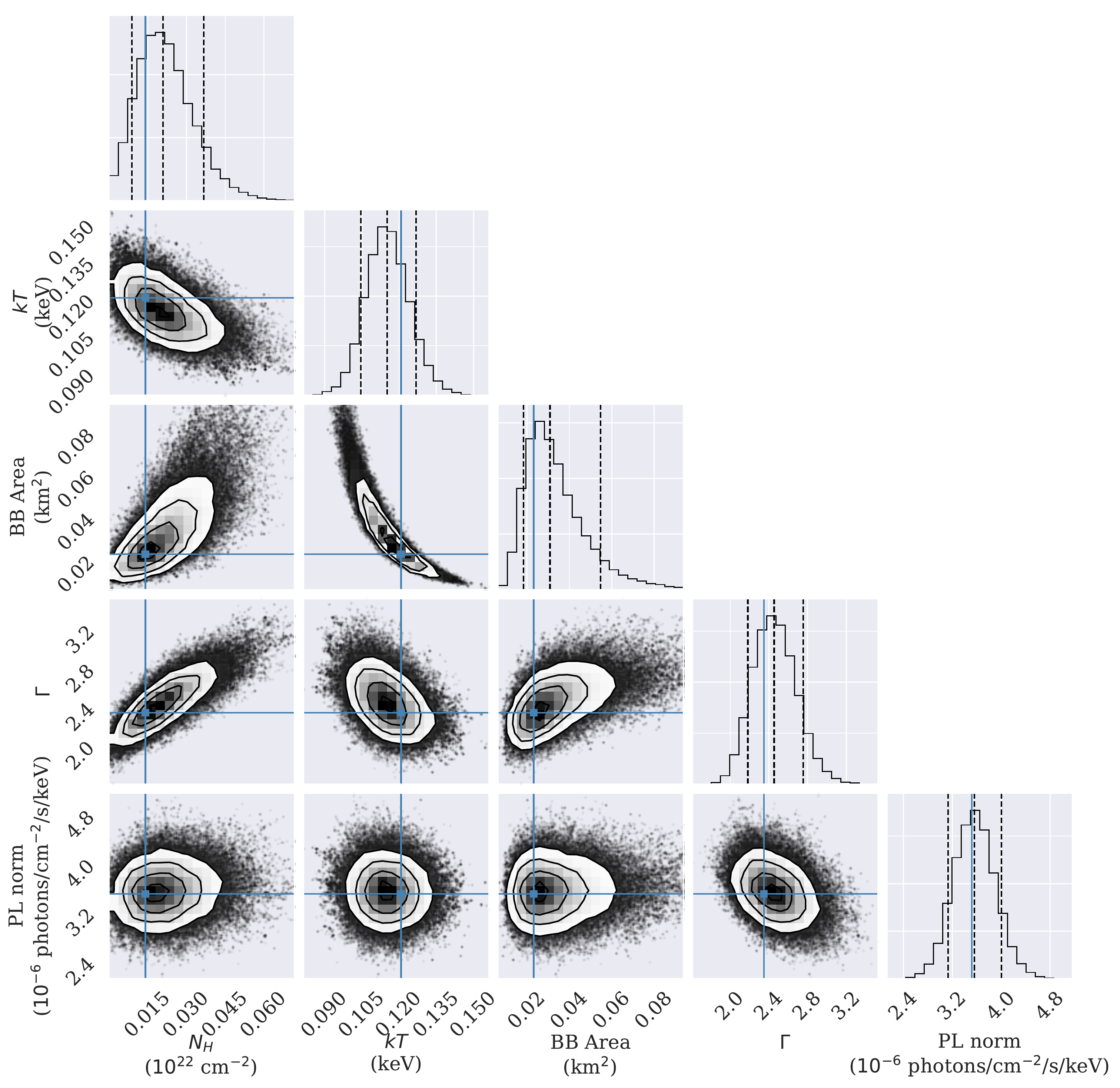}
\caption{\label{fig:PhRcorner} 1D marginalized distributions and 2D marginalized joint-plots between the parameters in the joint BB and PL fit to the phase-separated spectra from phase ranges $0.2-0.7$ and $0.7-0.2$, respectively.
The parameters corresponding to the highest posterior probability are shown with blue lines.
The dashed lines in the 1D distributions show the 16, 50, and 84 percentile limits.
The contour lines in the 2D joint plots show the $1,2,\text{and}\,3-\sigma$ bi-variate uncertainties for pairs of parameters.}
\end{figure*}

Single-component models, either thermal or non-thermal, are sufficient to explain the spectra.
We obtained the best fit (Figure \ref{fig:J0108PhRSpectrum}) and constraints (Table \ref{tab:spectralfit}; Figure \ref{fig:PhRcorner}) on the model parameters when simultaneously fitting a BB for S1 and a PL for S2, while using a shared ISM absorption parameter ($N_{\rm H}$ tied).
The resulting constraint on $N_{\rm H}$ is stronger due to the improved statistics from using data from both the spectra than otherwise obtained from individual spectral fits.
This also better constrains the emission model parameters.
A PL fit to S1 produces unrealistically steep photon index $\Gamma\approx5$ and results in a bad overall fit ($\chi_\nu^2 = 1.5$), while a BB fit to S2 is overall bad ($\chi_\nu^2 = 1.7$) with large systematic residuals.

BB emission flux is directly proportional to the emitting area and the spectral shape is uniquely determined by the temperature parameter.
The polar cap area on a NS of radius $R_{\rm NS}$ and spin period $P$ for an spin-aligned magnetic dipole is
\begin{equation}
A_{\rm d,pc} = 2\pi^2R_{\rm NS}^3/(cP)
\end{equation}
where, $c$ is the speed of light \citep{Sturrock1971}.
For J0108, assuming $R_{\rm NS}=10$ km and mass of the NS $M_{\rm NS}=1.4\,M_{\sun}$, $A_{\rm d,pc} = 0.0815\,{\rm km}^2$, which for an observer at infinity is $A_{\rm d,pc}^\infty = A_{\rm d,pc}/g_{\rm r}^2 = 0.139\,{\rm km}^2$, where, $g_{\rm r} = \sqrt{1-2.952\,M_{\rm NS}/R_{\rm NS}}$ is the gravitational redshift parameter.
Through Bayesian analysis, we obtain a representative sample of the probability distribution function (PDF) for the emitting area (the marginalized posterior probability distribution conditional on the observed spectrum) assuming the nominal distance estimate of 210 pc.
The median value $A_{\rm d,pc}^\infty = 0.032_{-0.014}^{+0.031}$ km$^2$, with uncertainties obtained from the 10-90 percentile interval.
We estimate the probability that the measured BB area is lower than the nominal value of $A_{\rm d,pc}^\infty$, $P(A < A_{\rm d,pc}^\infty) = 99.8$\%.
We propagate the pulsar's rather large and asymmetric distance uncertainties, used in the BB area estimates, by sampling a distribution of the form
\begin{align}
    \begin{split}
    P(d)\;=&\;H(D-d)\;\exp\left[-0.5\times\left(\frac{d-D}{D_{\rm min}}\right)^2\right]\\
    +&\;H(d-D)\;\exp\left[-0.5\times\left(\frac{d-D}{D_{\rm max}}\right)^2\right],
    \end{split} \\
    H(x)\;=&\;\begin{cases}
                0, & \text{if $x<0$}.\\
                0.5, & \text{if $x=0$}.\\
                1, & \text{if $x>0$}.
        \end{cases}
\end{align}
where, $D_{\rm min}, D, D_{\rm max}$ are 160, 210, and 310 pc, respectively \citep{Verbiest2012}.
The probability that the measured BB area is lower than $A_{\rm d,pc}^\infty$, after including the distance uncertainty, is $P(A < A_{\rm d,pc}^\infty) = 86$\%.
The ratio of the measured hot spot area to $A_{\rm d,pc}^\infty$ is $0.48_{-0.30}^{+0.78}$, with uncertainties quoted at 90\% probable limits about the median value of the distribution.

We tried to constrain the phase of the thermal peak by comparing thermal emission component contribution in the two spectra S1 and S2.
We simultaneously fit a BB model to S1 and a PL+BB model to S2 while keeping the parameters $n_H$ and $kT$ tied-up between the two spectra (Table \ref{tab:spectralfit}).
The area parameter estimates in the two phase ranges show larger area values for spectrum S1 compared to S2 (Figure \ref{fig:J0108PhRSpectrum}).
We calculate a 98\% probability that the BB area is larger in the $0.2-0.7$ phase range when compared to that in the $0.7-0.2$ phase range.

\subsection{Beamed Thermal Emission Models}
A BB is not the only thermal emission model that can explain the neutron star surface radiation.
Local magnetic field structures can produce beamed radiation by introducing anisotropy in the conductivity of the stellar crust, radiation transfer in a layer of NS atmosphere, or charged particle motion that Comptonize the radiation.
To compare the effect of beaming on observed emission areas, we fit neutron star atmosphere models (NSA and NSMAXG) to the soft spectrum extracted in the $0.2-0.7$ phase range.

\begin{figure*}
\captionsetup[subfigure]{labelformat=empty}
\subfloat[]{\label{fig:J0108BB-NSAa}\includegraphics[width = 0.48\textwidth]{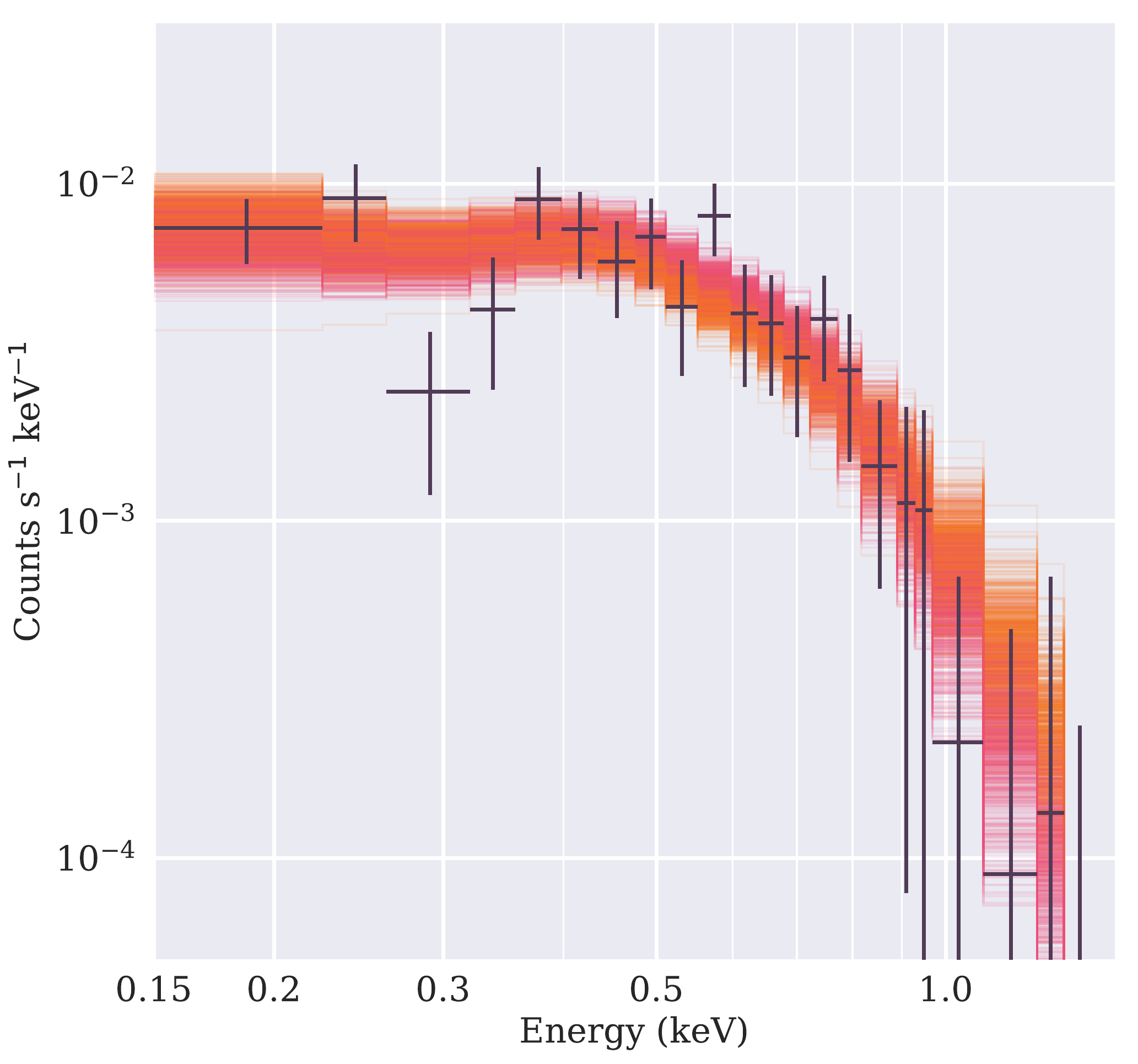}}
\subfloat[]{\label{fig:J0108BB-NSAb}\includegraphics[width = 0.48\textwidth]{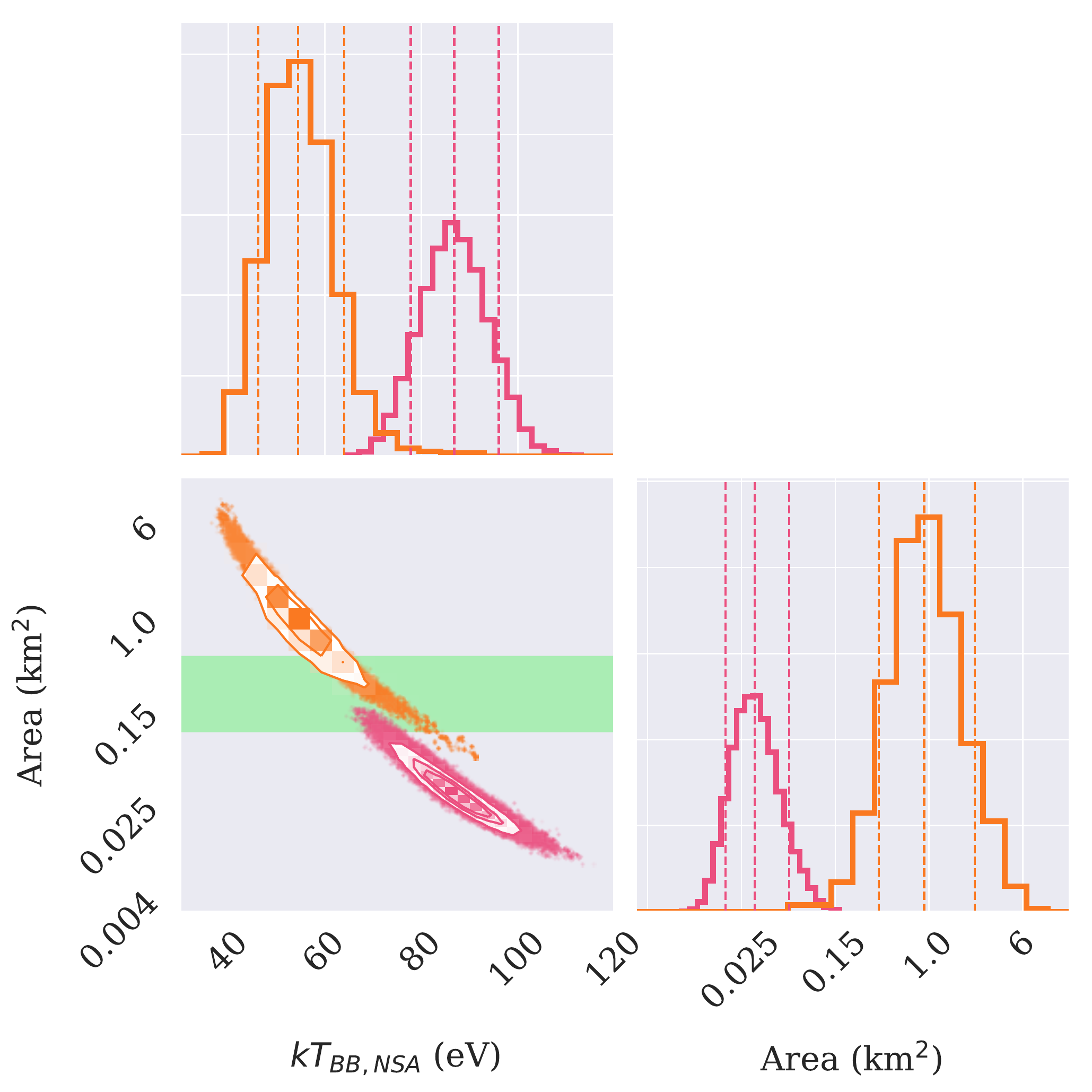}}
\caption{Top: BB (Red) and NSA (Orange) models (left) and spectra(right) that fit the observed spectrum (Brown errorbars) in the $0.2-0.7$ phase range, sampled from the posterior distribution.
Bottom: Comparison of the model parameters emission area and effective temperature ($kT$) for BB and NSA models using their posterior distributions.
The green band shows the uncertainty in the NS PC area when NS mass ($M_{\rm NS}=1.49\pm0.16\,M_{\sun}$) and radius ($R_{\rm NS}=8-16$ km) uncertainties are propagated.
}
\label{J0108BB-NSA}
\end{figure*}

We fit for the effective temperature and the normalization parameters in the NSA model while freezing the NS mass at $1.4\,M_{\sun}$, NS radius at 10 km, and the surface magnetic field strength at $10^{12}$ G.
The overall fit of the PL+NSA model to the spectra (NSA to S1 and PL to S2) is comparable to the PL+BB fit ($\chi_\nu^2 = 1.2$).
The $10-90$\% parameter range about the median for the effective temperature $T_{\rm eff} = 9.0_{-1.6}^{+2.1}\times10^5$ K and the emission area to NS area ratio $A_{\rm ratio}=2.6_{-1.7}^{+5.3}\times10^{-4}$ ($A_{\rm pc}/A_{\rm NS}=6.3\times10^{-5}$).
Comparing the fit parameters --- temperature and emission area --- between the two models, we see that the spectrum is unable to differentiate between a set of BB fits and an equivalent set of NSA fits with colder temperatures and larger emitting areas (Figure \ref{J0108BB-NSA}).
The comparison shows that spectra from a beamed, reprocessed thermal emission model such as from a NS atmosphere, cannot be statistically distinguished from a BB.
In this instance, fitting BB to a thermal spectrum processed by a NS atmosphere would lead to under-estimated emission areas and over-estimated temperatures.

\subsection{Absorption-like features in the thermal component}
The spectrum in the phase-range $0.2-0.7$ shows an absorption-like feature at approximately 0.3 keV.
The feature, seen as un-modeled residuals in thermal BB or NSA model fits, is hard to explain with purely instrumental systematics.
The $\sim40\pm20$\% residuals are significantly greater than the $3\sigma$ upper-limit on the systematic uncertainty in the effective area of $4\%$ estimated for the EPIC-pn detector\footnote{\url{http://xmm2.esac.esa.int/docs/documents/CAL-TN-0018.pdf}}.
Moreover, the feature is only seen in the spectrum from one-half of the rotational phase (Figure \ref{fig:J0108PhRSpectrum}).

We refer only to the model components fit to spectrum S1 below, even though we always simultaneously fit a PL to S2.
Introducing an absorption line modeled with a Gaussian profile (\texttt{Gabs} in XSPEC) to the NSA model, we see a marginal improvement in the test statistic ($\chi_\nu^2=1.1$ for Gabs$\times$NSA versus $\chi_\nu^2=1.2$ for NSA).
We estimate the central energy of absorption $E_{\rm c}=0.33_{-0.08}^{+0.05}$ keV with equivalent width $EW=0.105\pm0.05$ for NSA continuum with $\log T_{\rm eff}=5.92_{-0.07}^{+0.05}$ and area ratio $A_{\rm ratio}=4.1_{-1.8}^{+6.8}\times10^{-4}$.

\noindent Note on statistical model selection:\\ The models (BB/NSA/NSMAXG)+PL and Gabs(NSA+PL), fit simultaneously to the phase-separated spectra, all produce statistically acceptable fits.
A measure such as the Deviance Information Criteria (DIC; \citealt{Spiegelhalter2002,gelman2003bayesian}) fails to statistically distinguish between the models, with the values lying in a narrow range of 64--71\footnote{A difference of $\geq10$ is typically considered strong evidence and $5-10$ may be considered as substantial evidence in favor of the model with the lower DIC.}.

\section{Thermal X-ray and Radio Phase Alignments}
\label{radem}
In the earlier sections, we have demonstrated that thermal emission mechanism cannot be uniquely identified as BB and consequently, the PC area can be very different depending on the emission model assumed.
Hence, X-ray spectra cannot unambiguously provide evidence for the surface multipolar magnetic field in J0108.

An alternative way of accessing the presence of multipolar magnetic fields is to look for phase offsets between the time-aligned radio and X-ray pulse profile.
The radio emission is known to arise a few hundred km above the neutron star surface.
Hence, if the thermal X-ray emission originates from regions on the surface with a multipolar magnetic field, the higher order multipoles fall off rapidly with height, and at the radio emission height, the magnetic field is purely dipolar.
Thus, in the presence of a multipolar magnetic field, the time-aligned thermal X-ray and radio profiles can show significant phase offsets.
In this section, for J0108, we first check the validity of the assumption that the radio emission arise from regions where the magnetic field is dipolar and then time-align the radio profile with the thermal X-ray peak to obtain the phase-offset.

\subsection{Radio Emission, profile, geometry, and Emission heights}

\begin{figure}
\centering
\includegraphics[trim=4.2cm 0 2.6cm 1cm,clip,width = 0.47\textwidth]{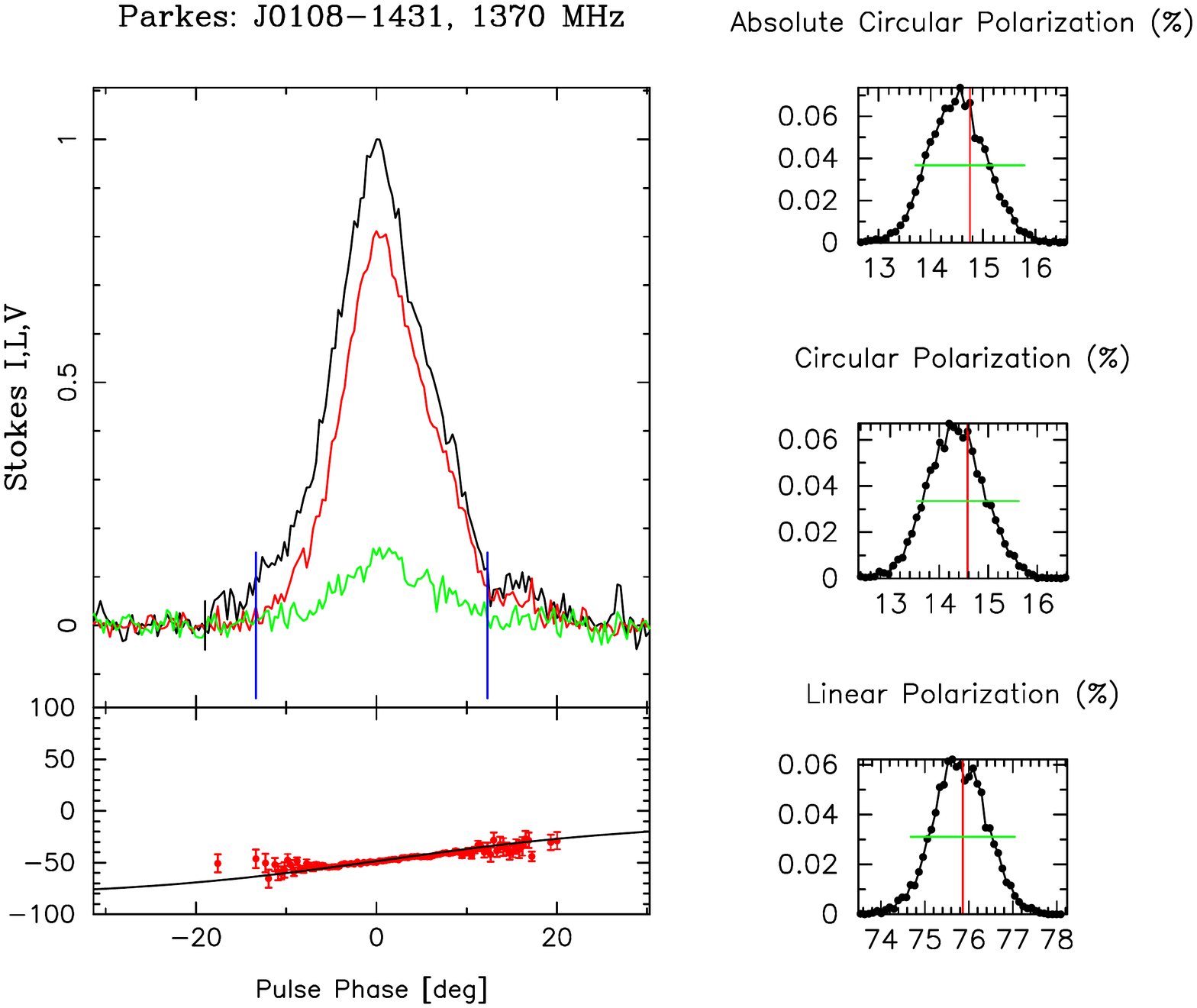}
\caption{\label{fig:radio}
The left plot shows the time-averaged polarisation properties of PSR J0108--1436 at 1370 MHz using the data from \protect\cite{Johnston2018}.
The top panel of the plot shows the total intensity (black line), linear polarization, $ L = \sqrt{U^2 + Q^2}$ (red line), and circular polarization $V$ (green line).
The two vertical blue lines near the trailing and leading edge of the profile correspond to 5 times the root mean squared (rms) estimate of the off-pulse total intensity.
The bottom panel shows the polarization position angle (PPA) $\psi = 0.5 \tan^{-1}(U/Q)$, and only points which are $\gtrsim5$ times the rms of the linear polarization are plotted.
The black line corresponds to the RVM fit for $\alpha = 7.5$, $\beta=6.2$ and the zero longitude correspond to steepest gradient point of the PPA traverse.
The three panels on the right show the distributions of the variation of average \%L, \%V, and $\mid V \mid$ (from bottom to top) that arises due to the noise in the off pulse region using the scheme discussed in \protect\cite{Mitra2016}.
The red line shows the median of the distribution and the green line shows the rms.
}
\end{figure}

Radio observations for PSR J0108--1431 exist between 400 MHz to 1.4 GHz and the average profile has a single component across these frequencies. 
The outer half power width of the profile at 430 MHz is 18$^{\circ}\pm1^{\circ}$ \citep{Lorimer1994} and that at 1370 MHz is $11\fdg5\pm0\fdg3$ \citep{Johnston2018}.
These values are consistent with the phenomenon of radius to frequency mapping (RFM) in pulsars wherein the pulse widths are seen to decrease with increasing frequency.
High-quality average polarization observations at 1.37 GHz by \cite{Johnston2018} reveal that the pulsar is highly polarized with about 76\% linear and 15\% circular polarization.
The polarization position angle (PPA) shows a smooth, flatter, and almost linear traverse across the pulse profile (see also Figure~\ref{fig:radio}, where the average profile
from the data from \cite{Johnston2018} is plotted).

Based on the available data, the Empirical Theory of pulsar emission (ET:I-VII, \citealt{Rankin1983I, Rankin1983II, Rankin1986III, Rankin1990IV, Radhakrishnan1990V, Rankin1993VI, Mitra2002VII}) may be used to infer the geometry of the system. 
The flat PPA traverse is consistent with the rotating vector model, according to which the electric field of the radio emission lies either in the plane or perpendicular to the dipolar magnetic field line planes.
As a result, the PPA traces the magnetic field planes, and near the region where the line of sight crosses the fiducial plane containing the rotation and the magnetic axis, a characteristic S-shaped curve is seen.
The shape of the S-curve is a function of the angle between the rotation axis and magnetic axis $\alpha$ and the angle between the magnetic axis and the observer's line of sight $\beta$.
The slope of the PPA is maximum near the profile center ($\mid d\Psi/d\phi \mid_{max}$) and is related to the geometrical angles as,
\begin{equation}
\sin(\alpha)/\sin(\beta) = \mid d\Psi/d\Phi \mid_{max}
\label{eq1}
\end{equation}
The value of the slope is estimated to be $\sim1.2$ for J0108.
In principle, the PPA traverse can be fitted to obtain the values of $\alpha$ and $\beta$, however, the fitted $\alpha$ and $\beta$ are highly correlated and they do not provide any meaningful values (\citealt{vonHoensbroech1997, Everett2001, Mitra2004}).

Instead, we can use the ET theory, according to which the pulsar radio emission beam is in the form of a core-cone structure, where, in the vicinity of the dipolar magnetic pole, there is a Gaussian-like core emission surrounded by two nested conal emission, namely the outer and inner cone.
The beam is roughly circular in shape having a beam radius $\rho$, which is related to $\alpha$, $\beta$, and pulse width $\phi$ as, 
\begin{equation}
\sin^2(\rho/2) = \sin(\alpha + \beta)\sin(\alpha)\sin^2(\phi/4)+\sin^2(\beta/2)
\label{eq2} 
\end{equation}
In ET:IV, it was found that two distinct values of $\rho$ exist at 1 GHz for the outer and inner conal emission, which are related to the pulsar period as $\rho_{outer} = 5.7\,P^{-0.5}$ and $\rho_{inner} = 4.3\,P^{-0.5}$.
Further, in ET:VII, it was found that RFM is associated with outer conal emission, and since RFM is seen in J0108, we will assume here that the observed emission is from outer conal emission.

We calculate $\rho_{outer} = 6.3^{\circ}$ using J0108's period $P = 0.807$ s.
Now we can insert the value of $\rho_{outer}$, and $\beta$ obtained in terms of $\alpha$ from eqn.$\,$(\ref{eq1}), in eqn.$\,$(\ref{eq2}) and use an iterative procedure to find appropriate values of $\alpha$ and $\beta$ that will reproduce the observed 1 GHz half-power pulse width.
Following this method we found $\alpha \sim 7.5^{\circ}$ and $\beta \sim 6.2^{\circ}$. 
From $\beta / \rho \sim 0.98$, we infer that the line of sight traverses the beam rather tangentially, and since the pulse profile has a single component, we can classify the pulsar to be a conal single as per the classification scheme of ET.
Assuming the neutron star radius to be 10 km, a star-centered dipolar structure for its magnetic field, and same emission height across the radio profile, we can compute the radio emission height $h = 10\,P\,(\rho/1.23)^2 \sim  211$ km.

Our analysis above suggests that the radio emission properties of PSR J0108--1431 are consistent with emission arising from regions of open dipolar magnetic field lines, a few hundred km above the stellar surface. 
Since the line of sight cuts the emission beam rather tangentially, as suggested by the large $\beta$ values, we classify the profile as a conal single. The peak of the
profile hence can be considered to lie in the fiducial plane containing the rotation and the dipolar magnetic axis.

\subsection{Thermal X-ray and radio profile offset}

In this section, we discuss the method to measure the offset between the thermal X-ray emission peak and the radio profile peak.
We select a few X-ray events near the phase of $0.15-2$ profile peak as our X-ray times-of-arrival (TOAs).
We produced radio TOAs from archival Parkes profile data made available through the CSIRO data access portal.
Obtaining a combined timing solution, while allowing a constant offset parameter between the radio and X-ray TOAs, allows us to determine the absolute profile peak offsets between the two energy bands.

We perform non-parametric bootstrap re-sampling \citep{FeigelsonBabu2012} of the original list of event phases to obtain 10000 samples of event phases in the $0.15-2$ keV range.
The 10000 KDE-smoothed profiles produced from these sampled event phases are expected to represent the same underlying distribution (the true pulse profile), varying only due to the statistical variance inherent in the data (observed profile).
The peak of the $0.15-2$ keV profile is estimated to be at $\tilde{\phi}_x=0.003(40)$, where, the value in the parenthesis represent the uncertainty at the $10-90$ percentile level from the quoted median value.
We use the barycentered arrival times in MJD of eight randomly selected events from the narrow $0.003\pm0.003$ phase range to produce the list of X-ray TOAs, assigning a 33 ms uncertainty (corresponding to $\sigma_\phi=0.04$) to each TOA.
Due to low S/N, neither the time sub-integrated data produce reliable profiles nor the integrated profile produces a good template.
Hence, the X-ray events near the original profile peak and the peak uncertainty estimated through bootstrap re-sampling together produce the most conservative TOAs for the given data.

Archival data from the Parkes telescope for J0108 is available as part of a regular monitoring campaign of pulsars with high energy emission \citep{Weltevrede2010}.
We obtained radio TOAs over a long baseline (April 2009 to September 2011 $\equiv$ MJD $54931-55821$) around the X-ray observation epoch (June 2011 $\equiv$ MJD 55728).
We used PSRCHIVE (\citealt{Hotan2004,vanStraten2012}) to clean and reduce the sub-banded and sub-integrated profile data and produce integrated profiles.
We created a noise-free template from a very high S/N profile, and finally, obtain the TOAs using the Fourier phase gradient algorithm \citep{Taylor1992}.

\begin{figure}
\captionsetup[subfigure]{labelformat=empty}
\subfloat[]{\label{fig:tempo0}\includegraphics[width = 0.48\textwidth]{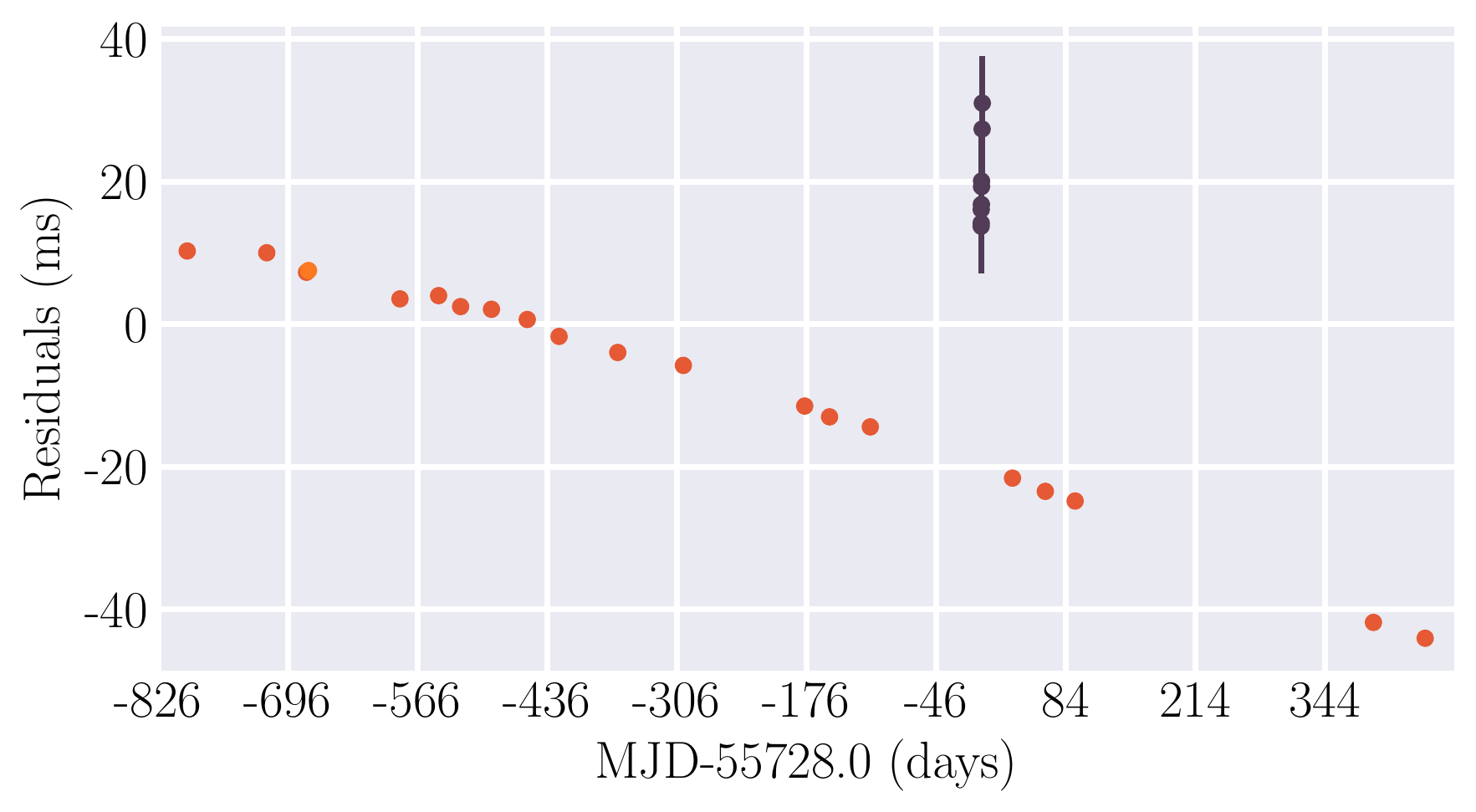}}\\[-6EX]
\subfloat[]{\label{fig:tempo2}\includegraphics[width = 0.48\textwidth]{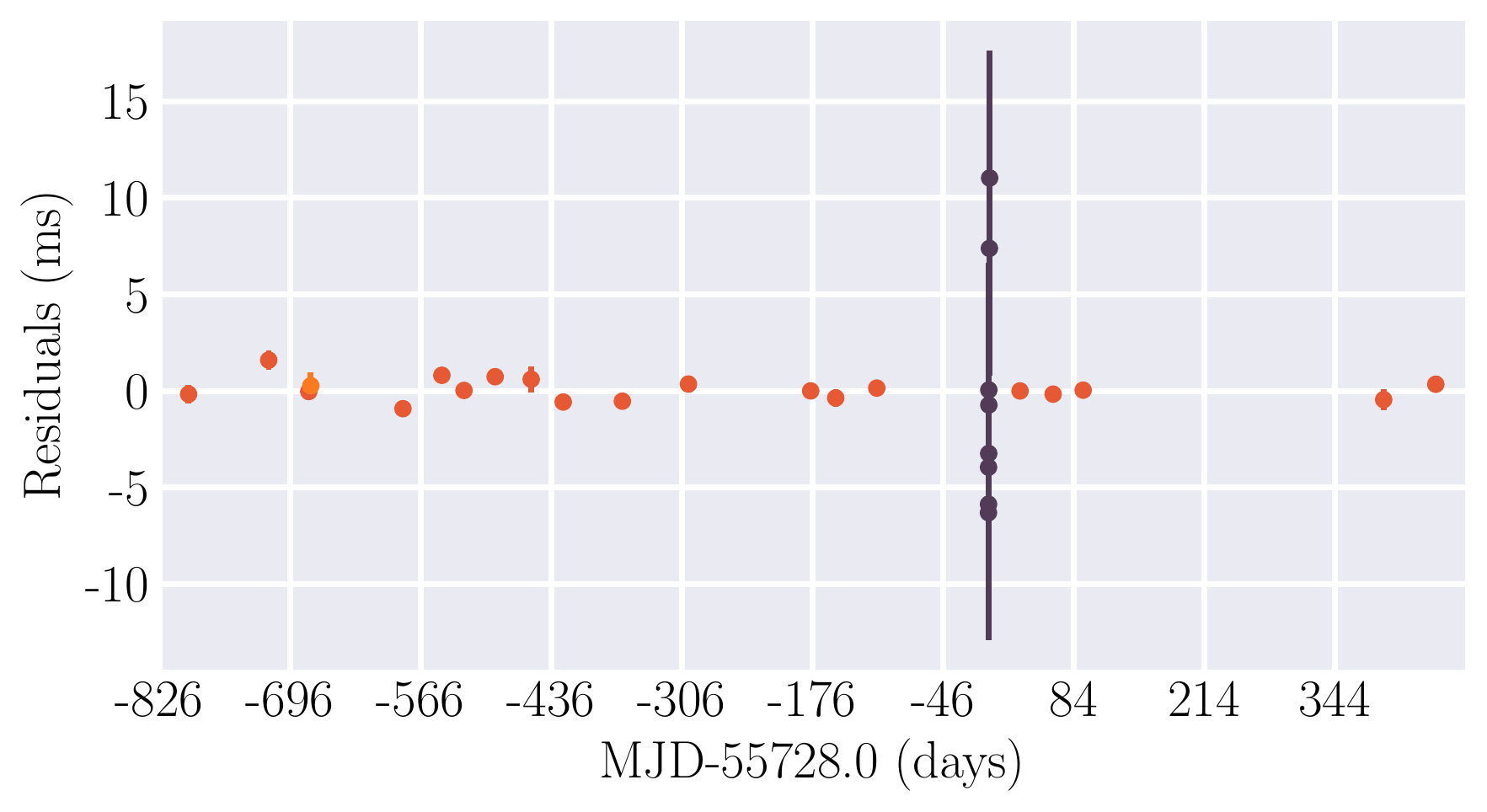}}
\caption{Timing and offset estimate from radio (orange) and X-ray (blue) TOAs. Pre-fit residuals (top) and post-fit residuals including the Radio-X-ray Offset parameter (bottom) are shown.
}
\label{fig:pulsartiming}
\end{figure}

\begin{figure}
\centering
\includegraphics[width = 0.47\textwidth]{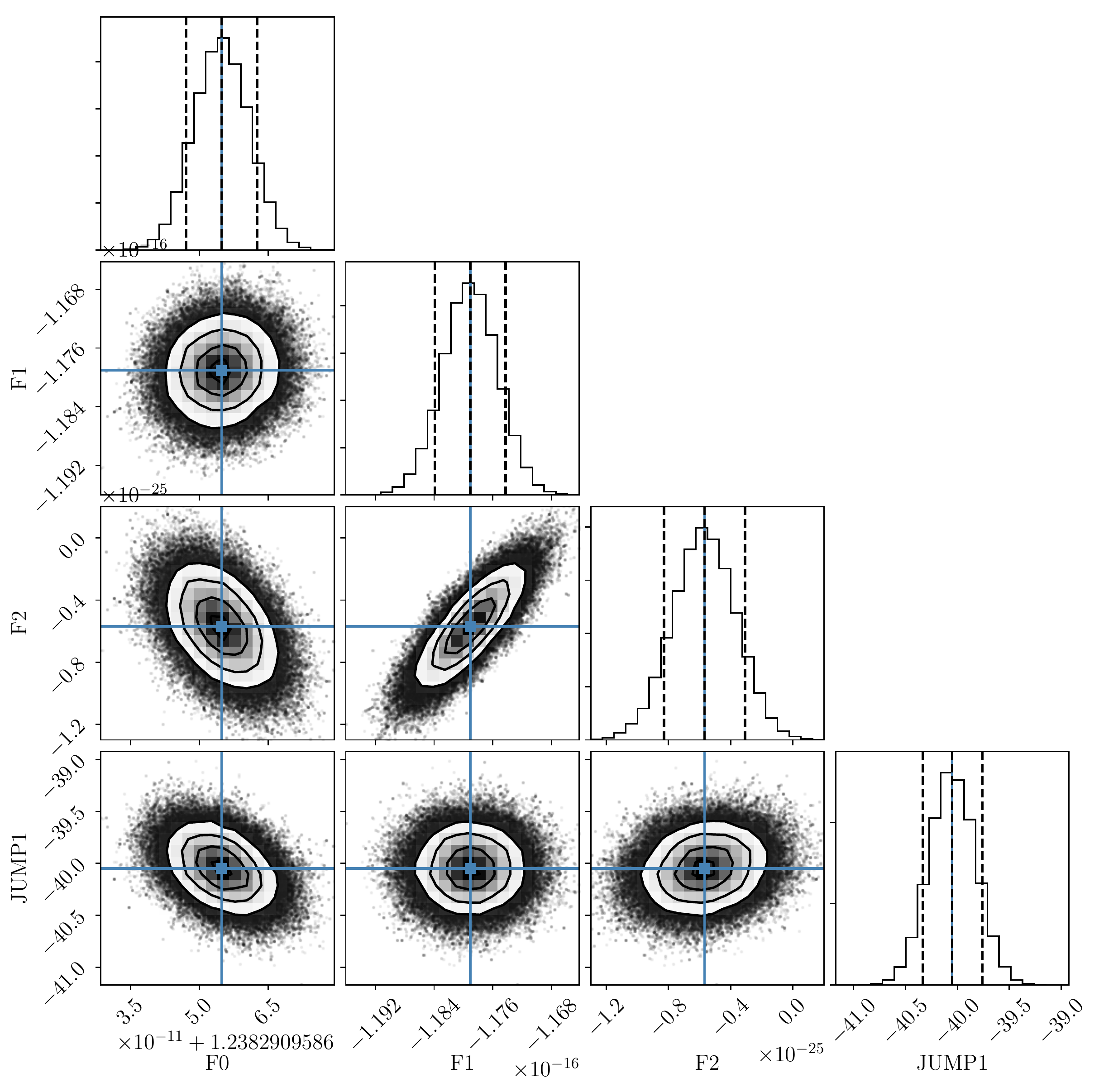}
\caption{\label{fig:T2corner} 1D marginalized distributions and 2D marginalized joint-plots between the pulsar rotational parameters: frequency in Hz (F0), frequency derivative in Hz s$^{-1}$, and frequency second derivative Hz s$^{-2}$.
Also shown is the posterior distribution of the X-ray profile offset (JUMP parameter in ms).
The parameters corresponding to the highest posterior probability are shown with blue lines.
The contour lines in the 2D joint plots show the $1,2,\text{and}\,3-\sigma$ bi-variate uncertainties for pairs of parameters.}
\end{figure}

We fit the TOAs using TEMPO2 \citep{Hobbs2006} through the python wrapper libstempo\footnote{\url{http://vallis.github.io/libstempo/}}.
Extrapolating the pulsar's timing solution to the epoch of X-ray observations require updates to only the rotation parameters: frequency ($F0$), frequency derivative ($F1$), and the frequency second derivative ($F2$).
We include the X-ray TOAs and introduce a constant offset parameter (JUMP) to find the radio-X-ray offset (Table \ref{tab:ephemeris}; Figures \ref{fig:pulsartiming} and \ref{fig:T2corner}).
The system delays in the Parkes radio data are of the order of a micro-second \citep{Manchester2013} which is orders of magnitude lower than our TOA uncertainties and hence are ignored.
The offset of the X-ray peak from the radio peak is estimated to be $\Delta\phi_{\rm x-r}=-0.04947(36)$ where the value in the parenthesis represent the uncertainty at the $10-90$ percentile level from the quoted median value.
Subtracting this from the phase of X-ray peak, the phase of the radio peak is estimated to be $\phi_{\rm r}=0.037_{-0.059}^{+0.041}$.

\begin{table*}
\centering
\caption{Timing Solution for PSR J0108--1431.}
\label{tab:ephemeris}
\begin{threeparttable}
\begin{tabular}{@{}llllll@{}}\toprule
Method & EPOCH & $F0$ & $F1$ & $F2$ & JUMP\\
      & (MJD) & (Hz) & ($\times10^{-16}$ s$^{-2}$) & ($\times10^{-27}$ s$^{-3}$) & (s)\\
\midrule
Original     & 50889 & 1.23829100810(3) & -1.1813(18) & -8.9(77) & ---\\
Expected\tnote{a} & 55728 & 1.23829095793322 & -1.2185     & -8.9 & ---\\\midrule
Fit Radio\tnote{b}   & 55728 & 1.2382909586474(77) & -1.1789(48) & -43(26) & ---\\
Fit X-ray\tnote{c}     & 55728 & 1.238290958651(12)  & -1.1822(67) & -53(38) & -0.03995(29)\\
\bottomrule
\end{tabular}
  \begin{tablenotes}
    \item[a] Extrapolated from known ephemeris.\\
    \item[b] Fit results with radio TOAs only.\\
    \item[c] X-ray and radio TOAs combined with a relative JUMP (offset) parameter.\\
  \end{tablenotes}
\end{threeparttable}
\end{table*}

\begin{figure}
\captionsetup[subfigure]{labelformat=empty}
\subfloat[]{\label{fig:sine1}\includegraphics[width = 0.48\textwidth]{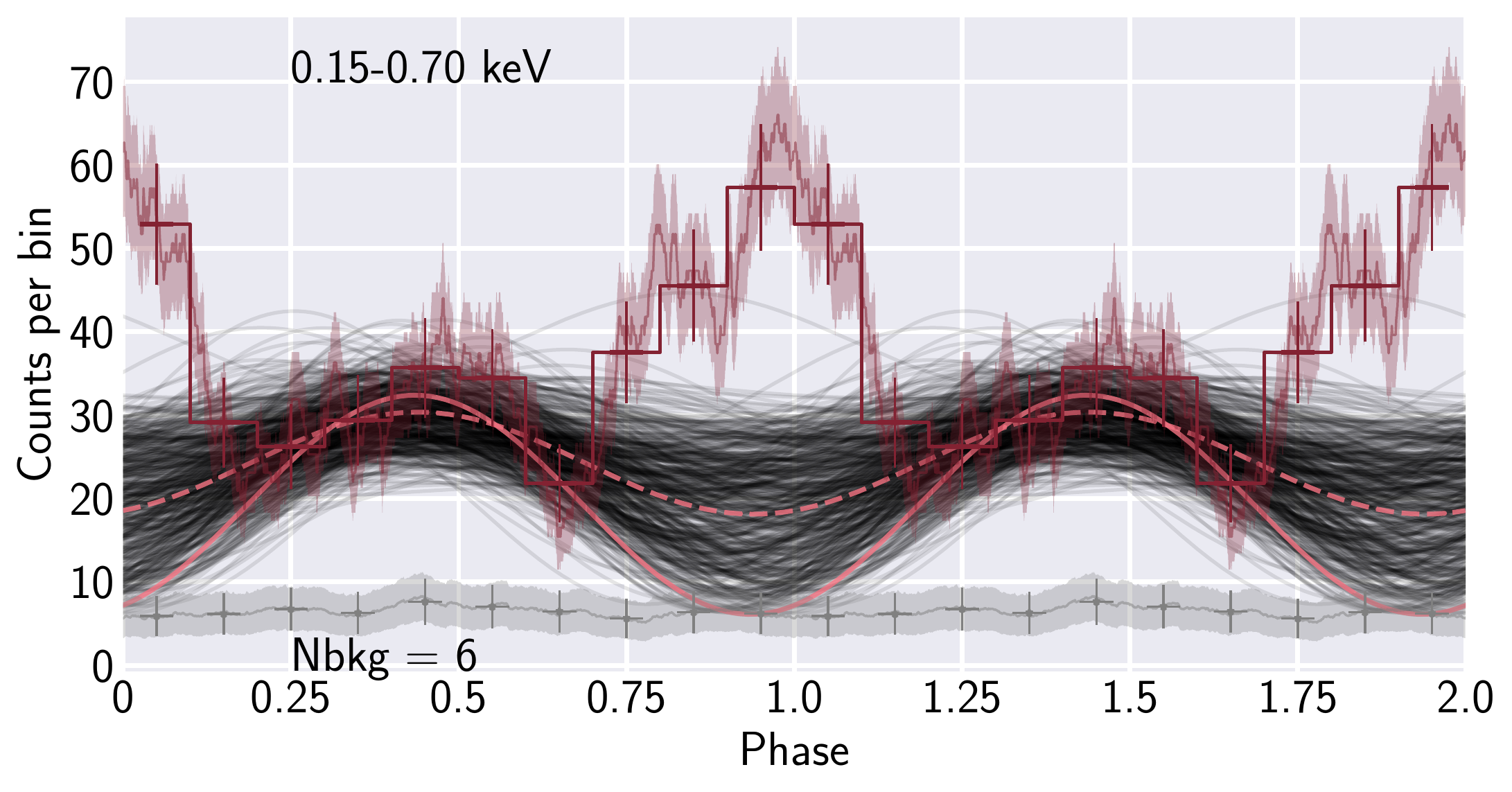}}\\
\caption{Sine curves (black) fit to the $0.15-0.7$ keV profile (red) in order to model BB emission.
The solid and dashed red Sine curves correspond to the maximum likelihood and median model parameters, respectively.
}
\label{fig:sinecurves}
\end{figure}

\begin{figure}
\centering
\captionsetup[subfigure]{labelformat=empty}
\subfloat[]{\label{fig:XRalign}\includegraphics[width = 0.48\textwidth]{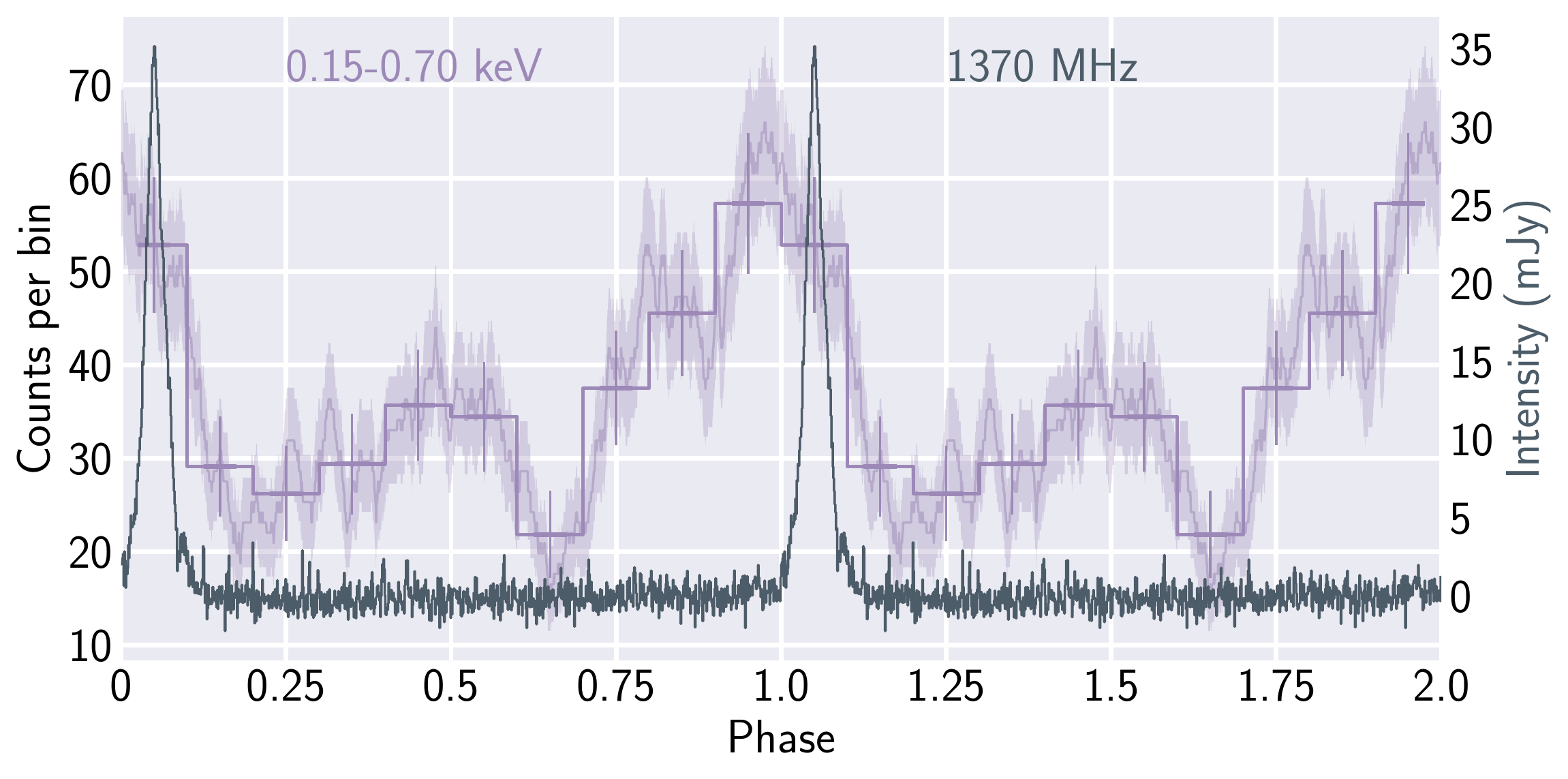}}\\[-5EX]
\subfloat[]{\label{fig:XRhist}\includegraphics[width = 0.44\textwidth]{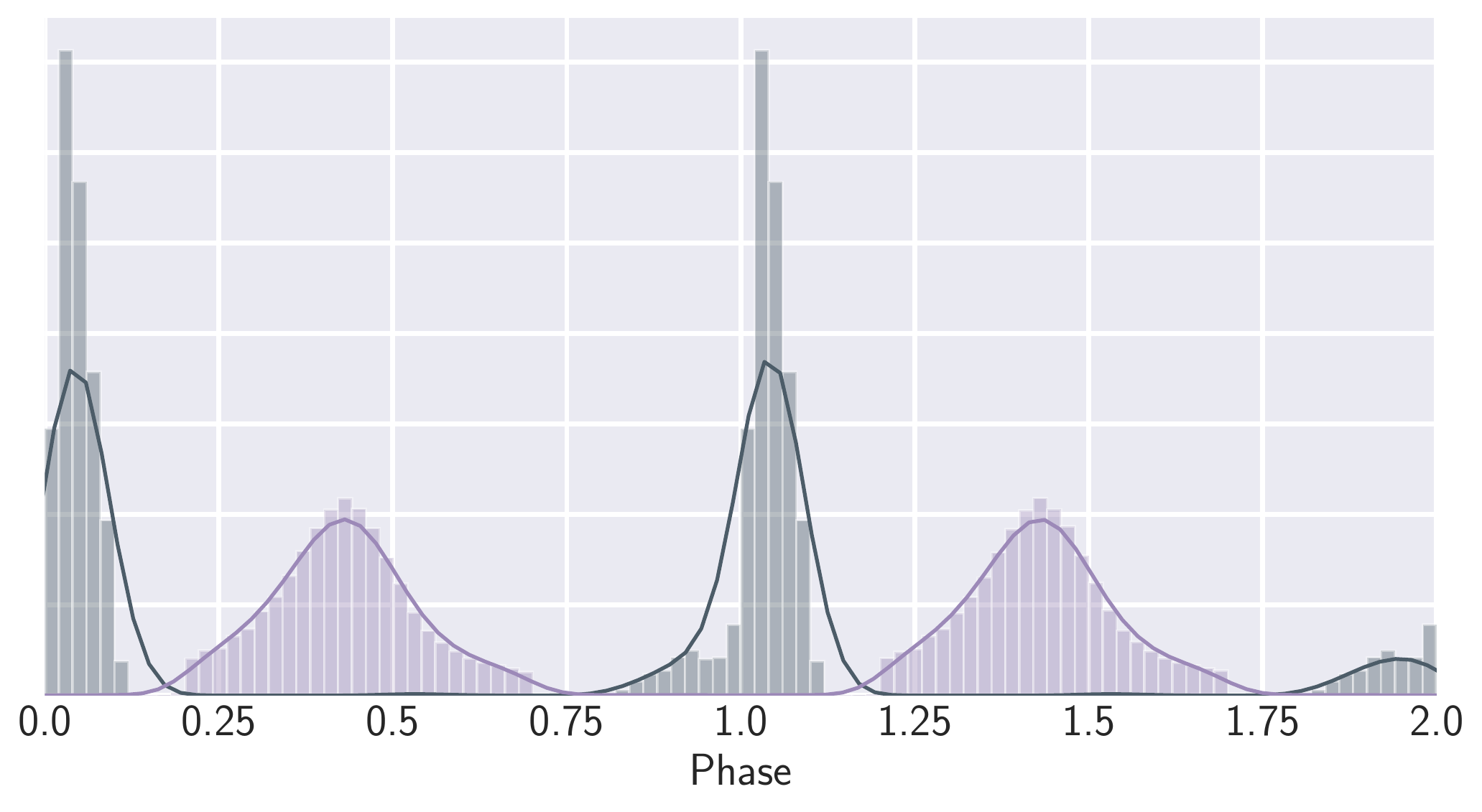}}
\caption{Top: Phase-aligned profiles radio (grey) and X-ray (violet) profiles.
Bottom: Histograms and KDE (solid curves) of the marginalized posterior distribution of sinusoidal peaks modeling the thermal emission peak (violet) and the Bootstrap estimated distribution for radio peak phase (grey).}
\label{fig:BBRpeaks}
\end{figure}

In order to obtain the peak of the thermal emission, we fit a sinusoidal model curve to the $0.15-0.7$ keV profile, in the $0.2-0.7$ phase range where the emission is thermal (Figure \ref{fig:sinecurves}).
The model curve is of the form
\begin{equation}
    f(x) = A0\,+\,A \times \sin\,[2\pi(x-\phi_0)]
\end{equation}
where, $A0, A$, and $\phi_0$ are the constant level, the amplitude of the Sine, and the zero phase, respectively.
We obtain a Bayesian 3 parameter fit, using MCMC to sample the posterior with 100 walkers taking 3000 steps after a burn-in of 500 steps.
Our prior for the zero-phase of the sine curve ($\phi_0$) is informed by the constraint on the thermal peak obtained from spectral fitting in Section \ref{subsec:BBspectral}.
The posterior distribution of the peak phase has 98\% of the values spread in the $0.2-0.7$ phase range and only 2\% outside.
Hence, we assign a probability of 0.98 that Sine curves peak in the $0.2-0.7$ phase range and 0.02 elsewhere.
The rest of the parameters are assigned uniform priors.
The peak phase of thermal emission is estimated to be $\tilde{\phi}_{\rm th}=0.43\pm0.14$ where, the uncertainties are at the $10-90$ percentile levels.
From the distribution for peak of the radio and thermal X-ray emission, we estimate the former to lead the latter by $\Delta\phi_{\rm r-th}=0.41_{-0.20}^{+0.13}$ (Figure \ref{fig:BBRpeaks}).
We also estimate a 99.7\% probability that the offset between the thermal X-ray peak and the radio peak is greater than 0.1.
This corresponds to a shift of $S \sim 2\,\pi\,\Delta\phi\,R_{\rm NS}\,\sin\alpha \gtrsim 820$ between the hotspot and the dipole axis.

\section{Discussions and Conclusions}\label{sec:discussion}

The need for a multipolar surface magnetic field has been recognized early on to explain significant pair multiplication above the pulsar polar cap \citep{RudermanSutherland1975}.
The presence of a strong multipolar magnetic field is often invoked to explain X-ray thermal emission from areas that are apparently smaller than the polar cap area for a dipolar surface magnetic field.
Recent compilations of polar cap radii from X-ray observed non-recycled pulsars have been used as strong observational evidence for a multipolar surface magnetic field (\citealt{Geppert2017, Rigoselli2018}).
We revisit the literature listed in these compilations and assess the robustness of their thermal polar cap area measurements.

We first identify systems with definite evidence of thermal emission.
Of the 18 pulsars in the lists, 9 of the pulsars do not require a thermal model component to explain the spectra (PSR J0108$-$1431; \citealt{Posselt2012}, PSR B1133+16; \citealt{Szary2017}, PSR B1451$-$68; \citealt{Pancrazi2012}, PSR B0628$-$28; \citealt{Tepedelenlioglu2005}, PSR B0834+06; \citealt{Gil2008}, PSR B1719$-$37; \citealt{Oosterbroek2004}, PSR B1929+10; \citealt{Misanovic2008}, PSR J2043+2740; \citealt{Becker2004}, PSR B2224+65; \citealt{Hui2012}).
For these pulsars, either the S/N of the data is too low to differentiate thermal contribution from the apparent single-component non-thermal emission, or a single PL component adequately explains the spectra.
In such cases, thermal polar cap areas are often estimated by over fitting the spectra with extra BB components, fixing the ISM absorption or non-thermal model parameters to obtain BB parameters, or from an upper-limit on the thermal component flux.
The PC area estimates from such systems are unreliable for scientific inference.

The estimates of the thermal emission area found in literature depends crucially on the assumption that the emission is a BB.
So, among the remaining 9 pulsars, none of them show definitive evidence for isotropic emission of a BB.
On the contrary, for pulsars with high S/N data, the pulse profiles are sharper and non-sinusoidal in the energy range where the polar cap emission dominates often with high pulsed fractions or increasing pulsed fraction with energy (PSR B0656+14; \citealt{DeLuca2005, Arumugasamy2018}, PSR B1055$-$52; \citealt{DeLuca2005}, PSR J0633+1746; \citealt{Kargaltsev2005}).
These are indicative of beamed (anisotropic) emission significantly deviating from a BB.
For rest of the pulsars, either no tests were performed and/or no evidence found to prove the existence of a BB (B0943+10; \citealt{Mereghetti2016}, PSR B1822$-$09; \citealt{Hermsen2017}, PSR B0355+54; \citealt{Klingler2016}, PSR B0114+58; \citealt{Rigoselli2018}, PSR J1740+1000; \citealt{Kargaltsev2012}, PSR J2021+4026 \citealt{Lin2013}).

Once the assumption of the BB emission model is relaxed, recovering the polar cap area from the observed thermal emission is not straight-forward.
In fact, as shown in this paper, fitting a BB to beamed emission from a neutron star atmosphere leads to under-estimate of the polar cap area.
BB area estimates found in literature also do not fully propagate the distance uncertainties which for DM based distances are usually not well quantified and for some parallax based measurements can be significant.
Hence, the observed PC radius estimates under the BB assumption are reliable only if there is clear evidence of thermal emission from the NS surface, followed by evidence that the thermal emission is a BB and not processed and beamed, and finally, the uncertainties in the distance and other component parameters are included.

We evaluate the evidence of a global dipolar magnetic field structure for the very old pulsar PSR J0108--1431 in this work.
Our results emphasize how low S/N data is incapable of uniquely determining the thermal emission model.
We answer the following questions:
\begin{enumerate}[label=(\Alph*)]

    \item Does the phase-integrated spectrum show clear evidence of a thermal component?\\
    A single $\Gamma\approx3$ PL component is sufficient to explain the phase-integrated spectrum.
    A single BB is not a good fit and two-component models (PL+BB or PL+NSA) over-fit the spectra, and hence are not required.
    We use the results of a PL+BB fit to get the rough energy ranges in which either of the two components could dominate.
    However, parameter estimates from such over-fit two-component models are unsuitable for any objective scientific inference.

    \item Is there evidence of multiple emission components (including thermal) in pulse profiles or phase-separated spectra?\\
    We see a strong indication of two distinct emission components separated in phase when comparing the profiles extracted in the $0.15-0.7$ keV and $0.7-2$ keV ranges.
    The softer component, seen in the phase range $0.2-0.7$, is only fit with a thermal component (BB or NSA) or an unrealistically steep, $\Gamma\approx5$ PL.
    The emission in the rest of the phase range ($0.7-0.2$) is best fit with a PL (BB or NSA models fit poorly) and hence predominantly non-thermal.
    
    \item Is the soft emission component fit only by a blackbody model?\\
    The soft spectrum in the $0.2-0.7$ phase range excludes a non-thermal description via a PL model.
    The spectrum can be explained with a 120 eV BB or a 78 eV NSA model emitting from a fraction, $3\times10^{-5}$ or $3\times10^{-4}$ of the NS surface, respectively.
    The statistics are insufficient to distinguish between the two models.
    
    \item Can the PC area be measured reliably?\\
    For an observer at infinity (far from the gravitational influence of the NS), the polar cap area for a star-centered, spin-aligned, global dipolar magnetic field rotating with a period of $0.808\,$s will appear to be $A^\infty_{\rm d,pc}=0.139$ km$^2$ (assuming $M_{\rm NS}=1.4\,M_{\sun}$ and $R_{\rm NS}=10$ km).
    The median BB emitting area estimate is a factor of 2 lower, with 86\% of the distribution below the dipolar value, after including the distance uncertainties.
    However, the median NSA emitting area estimate is a factor of 4 higher, with 80\% of the distribution above the dipolar value.
    Since the two models are statistically indistinguishable, we conclude that the area of the polar cap cannot be measured reliably with the current models.
    
    \item Are there any additional results from the X-ray analysis?\\
    We found an absorption-like feature in the spectrum extracted from half of the rotational phase (thermal continuum in the $0.2-0.7$ phase range).
    The feature cannot be explained by instrumental systematics or ISM absorption because it is absent in the spectrum from the rest of the phase.
    We fit the spectral feature with a Gaussian optical depth profile to obtain central absorption energy $E_{\rm c}=0.33_{-0.08}^{+0.05}$ keV with equivalent width $EW=0.105\pm0.05$.
    The absorption energy corresponds to local magnetic field strengths of $(2.2-3.3)\times10^{10}$ G or $(4-6)\times10^{13}$ G assuming electron or proton cyclotron absorption, respectively.
    This could be evidence for proton cyclotron absorption in the NS atmosphere with surface magnetic field strength exceeding $\sim10^{13}$ G.
    Such features with multiple harmonics and/or phase-dependent variation have been confirmed in high magnetic field pulsars (e.g., \citealt{Sanwal2002}; \citealt{DeLuca2004}; \citealt{Tiengo2013}; \citealt{Borghese2015}) where they are attributed to cyclotron absorption by protons confined in multipolar magnetic field structures close to the NS surface.
    \item What are the constraints on the coherent radio emission region?\\
    In section ~\ref{radem} we have demonstrated that the coherent radio emission from J0108 is consistent with the model of emission arising from a circular beam with underlying open dipolar magnetic field line region, at about few hundred km above the neutron star surface.
    We also infer that the pulsar is an almost aligned rotator, with the observer line-of-sight cutting the emission beam tangentially.
    For a star centered static magnetic dipole, the thermal hot spot and the radio emission region should be aligned in time.
    However, for the rotating case at an emission height of $\sim211$ km, aberration and retardation effects (\citealt{Blaskiewicz1991,Dyks2004}) advance the radio emission phase by $\Delta \phi \sim 0.011$, while magnetic sweepback \citep{DyksHard2004} introduces a lag $\Delta \phi \sim 0.015$ with respect to the phase of thermal hotspot emission.
    Hence, the radio peak is expected to lag the X-ray peak by $\Delta\Phi\sim0.004$ for a rotating dipolar magnetic field.

    \item Is the inferred surface thermal peak and radio peak offset for J0108 consistent with a star-centered dipole?\\
    We estimate an offset $\Delta\phi\approx0.4$ between the X-ray thermal peak and the leading radio peak, with a 99.7\% probability that the offset is greater than 0.1.
    Thus, the dislocation of the thermal polar cap with respect to the magnetic dipolar axis is $S \sim 2\,\pi\,\Delta\phi\,R_{\rm NS}\,\sin\alpha \gtrsim 0.8$ km, assuming $R_{\rm NS}=10$ km (see also \citealt{Szary2017}).
    An offset of such magnitude cannot be explained by a star-centered global dipolar magnetic field configuration.
    Hence, in the case of J0108, the measured X-ray thermal and radio emission offset provides a strong evidence for multipolar surface magnetic fields.
\end{enumerate}

We choose PSR J0108--1431 as a prototypical system to show how the prevailing method of determining PC area to infer the presence of a multipolar surface magnetic field is unreliable and how the radio and X-ray thermal emission offset method provides a better alternative.
In follow-up work, we plan to extend such analyses to the remaining sample of radio RPPs with thermal PC emission and determine whether strong evidence for multipolar magnetic fields can be found with the currently available data.

\section*{Acknowledgements}
We appreciate the ATNF and CSIRO Data Access Portal for making such high-quality data publicly available.
We thank the \xmm help-desk for important clarifications regarding the \xmm data analysis.
We also thank Bettina Posselt, Oleg Kargaltsev, George G. Pavlov, and Joanna Rankin for the useful discussions.
We greatly appreciate Dr. Sebastien Guillot for carefully reviewing our manuscript and suggesting essential improvements.
DM acknowledges funding from the grant ``Indo-French Centre for the Promotion of Advanced Research - CEFIPRA''.

\bibliographystyle{mnras}
\bibliography{references}


\bsp	
\label{lastpage}
\end{document}